\documentclass[twocolumn,apjl]{emulateapj} 
\usepackage{times}

\shorttitle{The Role of Starburst-AGN Composites in LIRG Mergers} 
\shortauthors{Yuan, Kewley and Sanders}
\journalinfo{Accepted by ApJ}
\submitted{Accepted by ApJ}

\newcommand {\ts}{\thinspace}
\newcommand{\oiii}{{[O\,{\sc iii}]}}
\newcommand{\nii}{{[N\,{\sc ii}]}}
\newcommand{\oi}{{[O\,{\sc i}]}}
\newcommand{\sii}{{[S\,{\sc ii}]}}

\def\OIIIHb{[{\ion{O}{3}}]/H$\beta$}
\def\NIIHa{[\ion{N}{2}]/H$\alpha$}
\def\SIIHa{[\ion{S}{2}]/H$\alpha$}
\def\OIHa{[\ion{O}{1}]/H$\alpha$}

\begin{document}

\title{The Role of Starburst-AGN composites in Luminous Infrared Galaxy Mergers:
 Insights from the New Optical Classification Scheme}

\author{T.-T. Yuan, L. J. Kewley and D. B. Sanders}

\affil{Institute for Astronomy, University of Hawaii, 2680 Woodlawn Drive, Honolulu, HI 96822}
\email{yuantt@ifa.hawaii.edu, kewley@ifa.hawaii.edu, sanders@ifa.hawaii.edu}

\begin{abstract}
We investigate the fraction of starbursts, starburst-AGN composites, Seyferts, and LINERs as a function of 
infrared luminosity ($L_{\rm IR}$) and merger progress for $\sim$500 infrared-selected galaxies.  Using the new optical 
classifications afforded by the extremely large data set of  the Sloan Digital Sky Survey,  we find that the 
fraction of LINERs in IR-selected samples is rare ($< 5$\%) compared with other spectral types. The lack 
of strong infrared emission in LINERs is consistent with recent optical studies suggesting that LINERs contain AGN 
with lower accretion rates than in Seyfert galaxies. Most previously classified infrared-luminous LINERs 
are classified as starburst-AGN composite galaxies in the new scheme. Starburst-AGN composites appear to ``bridge" the spectral evolution from starburst to 
AGN in ULIRGs.  The relative strength of the AGN versus starburst activity 
 shows a significant increase at high infrared luminosity.
In ULIRGs ($L_{\rm IR}>10^{12}L_{\odot}$), starburst-AGN composite galaxies dominate at early $-$ intermediate stages of the merger, and AGN galaxies dominate during the final merger stages. Our results are consistent with models for IR-luminous galaxies where mergers of gas-rich spirals fuel both starburst and AGN, and where the AGN becomes increasingly dominant during the final merger stages of the most luminous infrared objects. 
\end{abstract}


\section{INTRODUCTION}

Luminous Infrared Galaxies (LIRGs: $L_{\rm IR}>10^{11}L_{\odot}$\footnote{$L_{\rm IR} \equiv L(8-1000 \mu{\rm m})$; \citep[see][]{sanders96}}) 
were first discovered in small numbers $\sim$4 decades ago
 \citep{low68,kleinmann70a,kleinmann70b,becklin71,becklin72,rieke72}.  The importance 
 of these objects to galaxy evolution was made more clear following the first all-sky 
 survey carried out by the {\it Infrared Astronomical Satellite} \citep[{\it IRAS}:][]{neugebauer84}.   
 \citet{soifer87} found that the space density of infrared (IR)-selected LIRGs in the local Universe ($z < 0.1$) 
 rivaled that of the most powerful optically-selected starburst and Seyfert galaxies at similar bolometric 
 luminosity, and that the most luminous objects $-$ ultraluminous infrared galaxies  (ULIRGs: $L_{\rm IR}>10^{12}L_{\odot}$) $-$ had 
 similar space densities and bolometric luminosities as optically-selected quasi-stellar objects (QSOs). 
 
There have been numerous studies related to the origin and evolution of U/LIRGs\footnote{Previous studies of 
the properties of infrared galaxies versus log($L_{\rm IR}/L_{\odot}$) often divide the galaxy samples into decade luminosity 
bins and use the terms moderate, luminous, ultraluminous, and hyperluminous to refer to the decade ranges 
10$-$10.99, 11$-$11.99, 12$-$12.99, and 13$-$13.99, respectively.  We follow this convention here, and use 
U/LIRGs when we wish to refer to all galaxies with log($L_{\rm IR}/L_{\odot}$)= 11$-$12.99 .}, and most  
now seem to agree that strong interactions and mergers of gas-rich galaxies are the trigger for the 
majority of the more luminous LIRGs \citep[see the review by][]{sanders96}. The merger fraction
increases with IR luminosity and approaches 100\% for samples of ULIRGs 
\citep[{\it e.g.},][]{sanders88a, kim95a, clements96, farrah01}.  In the complete sample of 
IRAS 1{\ts}Jy ULIRGs by \citet{kim95a}, 117 out of 118 galaxies show strong
 signs of tidal interaction \citep{veilleux02}.  

There is less consensus on the nature of the power source of U/LIRGs.   It is clear  
that the IR luminosity in U/LIRGs can derive from dust reprocessing  
of either extreme starburst activity, Active Galactic Nuclei (AGN), or a combination of the two.
Studies of moderate to large samples of U/LIRGs \citep[{\it e.g.},][]{kim95a,veilleux95,goldader95,genzel98} 
indicate that the dominant power source in lower luminosity LIRGs is an extended 
starburst, and that an AGN often makes an increasing contribution to the bolometric 
luminosity in the more luminous sources with obvious energetic point-like nuclei.   
However, different studies of the same objects disagree on the relative contributions 
of starburst and AGN activity to the bolometric luminosity, in particular for the ULIRGs where 
the dominant energy source powering their extremely luminous and compact nuclear cores continues 
to be the subject of intense debate \citep[c.f.,][]{joseph99,sanders99}.  Although numerous studies at 
various wavelengths continue to be carried out to determine the energy source
of ULIRGs \citep[{\it e.g.},][]{tran01,farrah03,farrah07,lip03,ptak03,ima07}, determining
the relative contribution of starbursts and AGN within individual galaxies is still difficult.

One of the commonly proposed merger scenarios for ULIRGs \citep[{\it e.g.},][]{sanders88a,kim95b,farrah01,lip03,dasyra06} 
is based on the \citet{toomre72} sequence in which two galaxies lose their mutual orbital energy and 
angular momentum to tidal features and/or an extended dark halo and coalesce into a single galaxy.   
Tidal interactions and associated shocks are thought to trigger star formation \citep[{\it e.g.},][]{bushouse87,
kennicutt87, liu95, barnes04} which heats the surrounding dust, producing strong far-infrared (FIR) radiation.   
The FIR radiation rises to an ultra-luminous IR stage powered by starbursts and/or dust-shrouded AGN.   
As starburst activity subsides, the merger finally evolves into an optically bright QSO.  
In this scenario, ULIRGs plausibly represent a dust-shrouded transition stage that leads to the formation of 
optical QSOs \citep[{\it e.g.},][]{sanders88b,dasyra06,kawa06,zauderer07}.

A key element in testing the above scenario is to clarify the power source behind the strong IR emission, 
and the relationship between this power source and the evolutionary stage of the interaction.  
 Comprehensive studies on large IR-selected samples are crucial to this analysis.
Notable examples of such samples are the IRAS Bright Galaxy Survey  
\citep[BGS: ][]{veilleux95}, the IRAS 1{\ts}Jy ULIRGs sample \citep[1{\ts}Jy ULIRGs: ][]{kim95a, kim98, veilleux02},
 and the Southern Warm Infrared Galaxy sample \citep[SW01: ][]{kewley01a}. 

Most previous studies use standard optical spectral diagnostic diagrams 
to classify the dominant power source in emission-line 
galaxies \citep[]{baldwin81,veilleux87}.  
These diagrams are based on four optical emission line ratios that are sensitive 
to the hardness of the ionizing radiation field.   
More recently, the Sloan Digital Sky Survey (SDSS) has revolutionized these classification schemes 
by revealing clearly formed branches of star-forming galaxies, Seyferts, 
and Low Ionization Narrow Emission-line Region Galaxies (LINERs) on the diagnostic diagrams, 
for the first time \citep[]{kewley06}.   
Kewley et al. shows that many galaxies previously classified as LINERs lie 
along a well-defined mixing branch from star-forming galaxies to Seyfert galaxies.

In light of this new classification scheme, 
we investigate the new spectral classification of IR galaxies as 
a function of IR luminosity and merger progress.  
We describe our sample selection and derived quantities in \S~\ref{sample}.  
The results are presented in \S~\ref{results}. 
We discuss the results in \S~\ref{discussion}
and summarize in \S~\ref{summary}.
For convenience of comparison with the old 1{\ts}Jy ULIRG analysis by \citet{veilleux99,veilleux02},
 we adopt a Hubble constant of  H$_{0}$ = 75~ km~s$^{-1}$
Mpc$^{-1}$, and $\Omega_{m} = 0.27$, $ \Omega_{\Lambda} = 0.73$ throughout the paper.

\section{SAMPLE SELECTION AND DERIVED QUANTITIES}\label{sample}

\subsection{Sample Selection}
We use three local samples of IR-selected galaxies: the 118 ULIRGs from the IRAS 1{\ts}Jy sample of ULIRGs \citep{kim98} 
(hereafter the 1{\ts}Jy ULIRGs sample), 104 of the highest luminosity objects from the IRAS Bright Galaxy Survey 
(hereafter the BGS sample), and the complete sample of 285 galaxies in the Southern Warm Infrared Galaxies sample \citep{kewley01a} 
(hereafter the SW01 sample). 

The complete 1{\ts}Jy sample was compiled by \citet{kim95a} and is described in detail in \citet{kim98}. 
The 1{\ts}Jy sample was selected from the IRAS Faint Source Catalog (FSC) with flux 
$F(60\mu{\rm m}) > 1${\ts}Jy at high Galactic latitude $|b|>30^\circ$, and declination $\delta>-40^\circ$.  
The sample contains 118 objects with redshift $z=0.02-0.27$ and  log{\ts}$(L_{\rm IR}/L_{\odot}) = 12.00 - 12.90$. 
\citet{veilleux99} published optical spectra for 108 of these objects at a resolution of 8.3 \AA. 
Their nuclear spectra were extracted using a window corresponding to a physical diameter of 4 kpc 
(for the three objects with $z>0.2$: IRAS 00397$-$1312, IRAS 12032+1707, and IRAS 23499+2434, 
a diameter of 8 kpc was used).  Typical uncertainties for the emission line ratios are 5\%$-$10\%.
A $R$- and $K^\prime$-band image atlas for the 1{\ts}Jy sample is given in \citet{kim02} and the analysis of 
the morphological properties was carried out by \citet{veilleux02}. 

 We also include 104 lower luminosity objects from the IRAS BGS 
  \citep{sanders95, veilleux95,soifer86,soifer87,soifer89}.  The BGS represent all extragalactic sources
 brighter than 5.24 Jy at 60$\mu$m, $|b|>5^\circ$.  \citet{kim95b} provide optical spectra for 114 of these 
 objects at a resolution of 8$-$10 \AA. A constant linear aperture of 2 kpc was used to extract the nuclear spectra. 
 The redshift range is  $z=0.0027-0.09$ with a median of 0.02.  Among the original 114 BGS objects in \citet{kim95b}, 
 a total of 10/114 are ULIRGs,  with 8 of the ULIRGs also included  in the 1{\ts}Jy sample.  We only use the 104 LIRGs   
  (log{\ts}$(L_{\rm IR}/L_{\odot}) < 12.0$) in the BGS sample.  {\it The main role of  the BGS sample in this study is to 
  supplement the 1{\ts}Jy ULIRG sample with lower luminosity objects, and to help create a larger non-ULIRGs 
  sample in \S~\ref{dagn_ns}.}   The final $L_{\rm IR}$ range is log{\ts}$(L_{\rm IR}/L_{\odot}) = 10.5 - 13.0$ 
  for the 1{\ts}Jy ULIRG and BGS samples combined. 

The SW01 sample was selected by \citet{kewley01a} from the catalog of \citet{strauss92}.  
 It consists of 285 IRAS galaxies with $F(60\mu{\rm m}) > 2.5${\ts}Jy at $|b|>15^\circ$, $\delta < 0^\circ$. 
\citet{kewley01a} applied the ``warm" color
criteria ($F_{60}/F_{25}<8$)  to ensure that the sample contains a high fraction of AGN.   SW01
has a wide coverage in IR luminosity, and is dominated by LIRGs: among the total 285 galaxies, 
277 galaxies have log{\ts}$(L_{\rm IR}/L_{\odot}) = 8.0 - 11.99$ and  8 are ULIRGs.
 \citet{kewley01a} took high-resolution spectra (30~km~s$^{-1}$ at H$\alpha$) 
 for 235 objects in the SW01 sample (the emission line intensity measurements are accurate to within 30\%). 
 The SW01 redshift limit is $z< 0.027$ for IR luminosities log{\ts}$(L_{\rm IR}/L_{\odot}) <11.0$ and $z< 0.067$ 
 for log{\ts}$(L_{\rm IR}/L_{\odot}) >11.0$.   Their nuclear spectra were extracted using an aperture corresponding 
 to 1{\ts}kpc at the redshift of each galaxy. 

For the BGS and SW01 samples, we use optical images from the Digitized Sky Survey (DSS) and 
$K_{\rm s}$-band images from 2MASS. We use the $K_{\rm s}$-band images to obtain the projected nuclear 
separation for the interacting galaxies in our samples and we use the $R$-band and other available optical 
band images for identification of tidal debris.

\subsection{Optical Classification}\label{class}

\begin{figure*}[!t]
\epsscale{1.}
\plotone{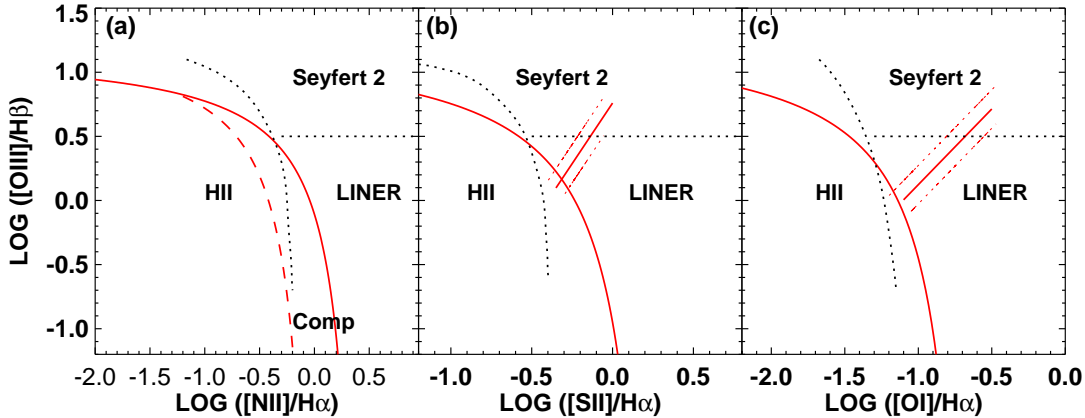}
\caption{Standard optical diagnostic diagrams showing the previous optical classification scheme  
by \citet{veilleux87} (black dotted lines) and the new classification scheme by \citet{kewley06}.  Star-forming galaxies form a tight abundance sequence on these diagnostic
diagrams.  The AGN branch begins at the metal-rich end of the star-forming galaxy sequence and extends towards the
upper right corner of these diagrams. 
Red solid curves are the theoretical  ``maximum starburst line" derived by \citet{kewley01b} as an upper limit for star-forming galaxies (see more descriptions in section 2.2);
the red dashed curve on the \nii\ diagram is the \citet{kauff03} semi-empirical lower boundary for the star-forming galaxies;
 the red lines (with the empirical error $\pm 0.1$~dex lines) on \sii\ and \oi\  diagrams are the empirical boundary lines 
  between Seyfert{\ts}2 galaxies and LINERs.  See \S~\ref{class} for more details.  The Kewley et al. (Ke06) scheme substantially changes the LINER boundaries, and includes a class for starburst-AGN composites 
(labeled Comp).   Note that in the \nii~$\lambda$6583/H$\alpha$ versus \oiii/H$\beta$ diagram (panel a), the VO87 scheme 
distinguishes between Seyfert{\ts}2 galaxies and LINERs, while the Ke06 classification scheme does not. }
\label{fig:class}
\end{figure*}

The classification of the dominant energy source in emission line galaxies using optical emission-line ratios was 
first proposed by \citet[][hereafter BPT]{baldwin81}.  BPT proposed the use of the \oiii~$\lambda5007/{\rm H}\beta$, 
\nii~$\lambda6583/{\rm H}\alpha$, and \oi~$\lambda6300/{\rm H}\alpha$ line ratios for spectral classification, taking 
advantage of the sensitivity of these line ratios to the hardness of the ionizing radiation field.  \citet{kennicutt84} 
and \citet{keel83} extended the initial set of classification ratios to include the \sii~$\lambda\lambda6716,6731/{\rm H}\alpha$\ 
line ratio which is also sensitive to the hardness of the ionizing radiation field and observable in the optical regime.  
To improve the optical classification, \citet{osterbrock85} and \citet[][hereafter VO87]{veilleux87} derived the first 
semi-empirical classification lines to be used with the standard optical diagnostic diagrams. Because of the pioneering 
work of \citet{baldwin81} and \citet{veilleux87}, the ``standard optical diagnostic diagrams" based on the 
\oiii/H$\beta$, \nii/H$\alpha$, \sii/H$\alpha$, and \oi/H$\alpha$ line ratios are commonly known as BPT or VO87 diagrams.   

Large samples or active galaxies reveal a tight abundance sequence for star-forming galaxies 
and an AGN sequence that begins at the metal-rich end of the star-forming abundance sequence and extends 
towards the upper-right corner of the diagnostic diagrams (i.e. towards large \oiii/H$\beta$, \nii/H$\alpha$, \sii/H$\alpha$, and \OIHa).

The first purely theoretical classification scheme was developed by \citet{kewley01b} 
(hereafter Ke01).  Ke01 used a combination of modern stellar population synthesis, photoionization, and shock 
models to derive a ``maximum starburst line" on the BPT diagrams.  
Galaxies that lie above this line can not be explained by any combination of starburst models and require a
dominant ($>50$\%) contribution from an AGN.   Galaxies that lie below the Ke01 line may include a non-dominant
(i.e. $<50$\%) contribution from an AGN.  

To obtain a more stringent sample of star-forming galaxies, \citet{kauff03} (hereafter Kau03) shifted the 
Ke01 line to form a semi-empirical upper boundary for the star-forming branch observed with the SDSS.  
The Kau03 line retains the shape of the Ke01 theoretical models, with a shift to enable classification of pure 
star-forming galaxies (i.e. 100\% star-formation dominated).  The combination of the Ke01 and Kau03 lines 
serves to separate pure star-forming galaxies, galaxies that are likely to contain both star-formation and 
AGN activity (composite galaxies), and galaxies that are dominated by their AGN. 

\citet[][hereafter Ke06]{kewley06} showed that AGN sequence forms two clear branches on 
the  \sii/H$\alpha$ and \oi/H$\alpha$ diagnostic diagrams.  These two branches were revealed with the 
large number of SDSS galaxies ($\sim$45,000);  these branches were not observed with 
 the smaller sample sizes ($\sim$200) that were used in previous studies (e.g., VO87 and Ke01).  
 Ke06 derived empirical boundary lines between Seyfert{\ts}2s and LINERs on the \sii/H$\alpha$ 
and \oi/H$\alpha$ diagrams based on the observed local minimum between the Seyfert and LINER branches. 
Seyfert and LINER galaxies defined in this way have significant differences in their host properties; LINERs are
older, more massive, less dusty, and less concentrated than Seyfert galaxies.  However at fixed accretion rate,
these differences disappear.   LINERs and Seyferts form a continuous sequence in Eddington rate from 
low to high Eddington rates, respectively.   Ke06 conclude that LINERs are AGN and that the dichotomy between 
Seyferts and LINERs is analogous to the high and low states observed in X-ray binary systems.  

As in X-ray binaries, LINERs have a harder ionizing radiation field and lower ionization parameter than Seyfert galaxies.
These characteristics make the \sii/H$\alpha$ and \oi/H$\alpha$ diagrams ideal for separating Seyferts and LINERs.
The \sii\ and \oi\ emission-lines are produced in the partially ionized zone at the edge 
of the nebula; this zone is large and extended for hard radiation 
fields.  Power-law AGN models from \citet{groves04} indicate that models with a hard 
radiation field and low ionization parameter are separated in the  \sii/H$\alpha$ and \oi/H$\alpha$ diagrams;
these ratios change by $\sim$0.7 dex as the power-law index is changed from -1.2 to -2.0.

Note that the \nii/H$\alpha$ ratio diagram can not be used to separate Seyferts and LINERs.  
The \nii/H$\alpha$ ratio is only weakly dependent on 
the hardness of the radiation field; log(\nii/H$\alpha$) only changes by 0.2 dex as the power-law 
index is varied from  -1.2 to -2.0 \citep{groves04}.  The \nii/H$\alpha$ ratio is much more strongly dependent on the metallicity of the nebular gas.  Metallicity differences among AGN host galaxies 
plus the weak dependence on hardness renders the \nii/H$\alpha$ diagram insensitive to the 
major differences between Seyfert and LINERs seen in the  \sii/H$\alpha$ and \oi/H$\alpha$ diagrams.

Ke06 estimated empirical errors ($\pm 0.1$~dex) 
for the Seyfert-LINER boundary by considering the positions of galaxies that remain Seyfert{\ts}2 or that remain LINER in 
both the \sii/H$\alpha$ and \oi/H$\alpha$ diagrams.  The class of galaxies that lie within $\pm 0.1$~dex  of the 
Seyfert{\ts}2/LINER line is uncertain.

 In Fig.~\ref{fig:class} we show the difference between the previous 
 VO87 classification scheme (black dotted lines) and the new Ke06 classification scheme (red solid and dashed lines).   
 Galaxies that were previously classified as LINERs may be either (a) true LINERs, (b) composite HII-AGN galaxies, 
 or (c) Seyfert{\ts}2 galaxies, or (d) high metallicity star-forming galaxies, according to the new classification scheme.  
 A substantial fraction ($\sim 1/3$) of ULIRGs and LIRGs have been previously classified as LINERs using the VO87 
 method \citep{veilleux95,veilleux99}.   Therefore, the application of  the Ke06 classification scheme may reveal new 
 insight into the power source behind IR-selected galaxies previously classified as LINERs. 

 In addition to the major change in LINER classification, the Ke06 scheme includes starburst-AGN composite
 galaxies as a separate class of objects.   The \nii/H$\alpha$ versus \oiii/H$\beta$ diagram is used 
 to classify composite galaxies.  (Composite galaxies lie between the red dashed and solid lines in Fig.~\ref{fig:class}a.)  
  The \nii/H$\alpha$ ratio is more sensitive to the presence of 
 a low-level AGN than the \sii/H$\alpha$, and \oi/H$\alpha$ line ratios because the  \nii/H$\alpha$ ratio is a 
 linear function of nebular metallicity until high metallicities where the \nii/H$\alpha$ reaches a plateau 
 at log(\nii/H$\alpha) \sim -0.5$ \citep{kewley02,denicolo02,pettini04}.   At this plateau, any AGN contribution 
 shifts the \nii/H$\alpha$ ratio above log(\nii/H$\alpha) > -0.5$.  An AGN contribution to low metallicity galaxies 
 is extremely rare \citep{groves06}.   
 
\begin{figure*}[!t]
\epsscale{1.34}
\plotone{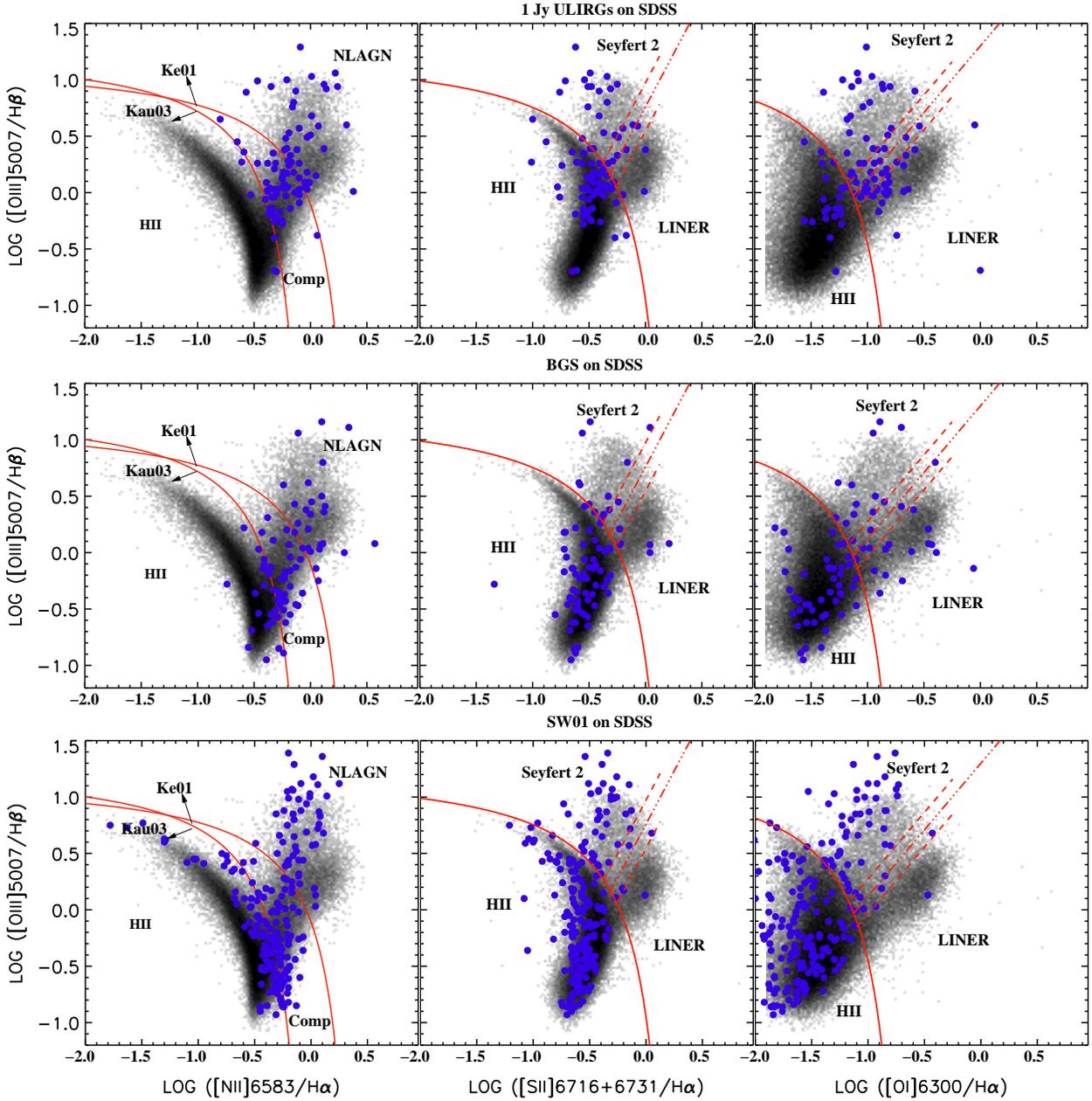}
\caption{\small Samples (blue filled circle) used in this work are shown on the BPT diagrams, over-plotted are the SDSS  
galaxies (black dots) from \citet{kewley06}.  
The curves and lines have the same meaning as in Fig.~\ref{fig:class}, i.e., 
upper red curves on all three BPT diagrams are the theoretical  ``maximum starburst line"; the lower red curve 
on the \nii\ diagram is the \citet{kauff03} semi-empirical lower boundary for star-forming galaxies, and
 the red lines (with the empirical error $\pm 0.1$~dex lines) in \sii\ and \oi\  diagrams are the empirical boundary lines between Seyferts and LINERs.  See the 
 text in  \S~\ref{class} for more details.
 From top to bottom, the samples are: 1{\ts}Jy ULIRGs, BGS, and SW01. 
The red lines are the new classification scheme \citet{kewley06} used to separate starburst (HII-region) galaxies, 
starburst-AGN composite galaxies, Seyfert{\ts}2, and LINERs.
 In the leftmost panels, NLAGN $\equiv$ narrow emission-line AGN (Seyfert{\ts}2 plus LINERs);   
Comp $\equiv$ starburst-AGN composites. }
\label{fig:sample}
\end{figure*}

We apply the new Ke06 classification scheme to our three samples to discriminate between 
star-forming galaxies  (or starburst/HII-region galaxies), Seyfert{\ts}2, LINERs, and starburst-AGN composites.  
In Fig.~\ref{fig:sample}, our three samples are shown in comparison with the SDSS galaxies used in 
\citet{kewley06} on the BPT diagrams with the new classification boundaries.    The Ke06 classification 
scheme that we use in this work is as follows: 

(1) Star-forming galaxies: lie below and to the left of Kau03 line on the \nii/H$\alpha$ diagram 
(e.g., Fig.~\ref{fig:sample} left column, the lower red solid line), and below and to the left of Ke01 theoretical 
lines in the \sii/H$\alpha$ and \oi/H$\alpha$ diagrams (e.g., Fig.~\ref{fig:sample} middle and right columns, the red solid lines).

(2) Starburst-AGN composites: lie above Kau03 line but below and to the left of Ke01 theoretical line in the \nii/H$\alpha$ 
diagram (Fig.~\ref{fig:sample} left column). 
 
(3) Seyfert{\ts}2 galaxies: lie above the Ke01 theoretical lines on all three BPT diagrams and also above the Seyfert-LINER 
boundary lines in the \sii/H$\alpha$ and \oi/H$\alpha$ diagrams (e.g., Fig.~\ref{fig:sample} middle and right  columns). 

(4) LINERs: lie above the Ke01 theoretical lines on all three BPT diagrams and below the Seyfert-LINERs boundary 
lines on the \sii/H$\alpha$ and \oi/H$\alpha$ diagrams. 

(5) Ambiguous galaxies: are those that are classified as one type of object in one or two diagrams and are 
classified as another type in the remaining diagram(s). 

(6) Seyfert{\ts}1 galaxies: {\it are not included on the BPT diagrams and are considered separately}.  They are characterized 
by their broad Balmer emission lines $-$ usually H$\alpha (FWHM) > 5\times10^{3}$~km~s$^{-1}$. Thus, galaxies classified as 
Seyfert{\ts}1 in previous studies remain classed as Seyfert{\ts}1 in our study.

The most stringent method for classification of galaxies using this scheme is to use all three diagnostic diagrams.
We refer to the use of all three diagnostic diagrams hereafter as the ``3-of-3" criterion.   The use of all three diagnostic diagrams allows for ambiguous galaxies to be classified separately.   For data sets with a significant fraction of 
unmeasurable or uncertain  \SIIHa\ or \OIHa\ ratios, an alternative ``2-of-3" criterion is often applied.  This method applies
the majority class, i.e. if 2 out of 3 diagnostic diagrams give one class, but the third diagram gives a different class or is unavailable, the consistent class of the 2 remaining diagnostic diagrams is assumed.  There is no ambiguous class if the 
2-of-3 criterion is applied.   Most previous studies on the optical classification of IR galaxies apply the 2-of-3 criterion 
\citep{veilleux95,veilleux99, veilleux02}. 
The \SIIHa\ classifications are uncertain for a  substantial portion ($\sim 16-17$\%)
of the BGS and 1 Jy ULIRG samples (i.e. the \SIIHa\ class lies within the 0.1~dex uncertainty line defined in Ke01).  
For comparison with previous work and to avoid contamination by uncertain \SIIHa\ classifications, we apply the  ``2-of-3" criterion to our samples.   We discuss our results in the context of the 3-of-3 criterion in the Appendix~\ref{ambissue}. 

In the 1 Jy ULIRG sample,the 2-of-3 classification scheme yields 8 (7.8\%) star-forming galaxies, 46 (44.7\%) starburst-AGN composites, 35 (33.9\%) Seyfert{\ts}2, 10 (9.7\%) Seyfert{\ts}1, and 4 (3.9\%) LINERs.  These classes include the 9 galaxies in the 1{\ts}Jy sample with double nuclei that have spectra taken for both nuclei.
Of these double nuclei galaxies, 4 galaxies have consistent classes for both nuclei and
are assigned composite (3/4) and Seyfert 2 (1/4) classes respectively.  The remaining 5 double nuclei galaxies have a composite nucleus plus either a starburst nucleus (4/5) or a Seyfert 2 nucleus (1/5).  Because their overall class is uncertain, we exclude these 5 double nuclei galaxies from our sample.  We note that our results are unchanged if we randomly assign these
double nuclei galaxies the class of either nucleus.

The BGS sample covers lower IR luminosities than the 1 Jy ULIRG sample and has a larger portion of star-forming galaxies.  The 2-of-3 classification scheme gives  30 (25.97\%) star-forming galaxies, 32 (41.56\%) starburst-AGN composites, 19 (24.67\%) Seyfert{\ts}2,  1 (1.3\%) Seyfert{\ts}1, and 5 (6.5\%) LINERs.  These classifications include 13 galaxies with double
nuclei in which both nuclei have consistent classes (7/13 star-forming and 6/13 composites).  We do not include an additional
5 double nuclei galaxies with differing classes for each nucleus.  Our results are not affected if we randomly assign these
double nuclei galaxies the class of either nucleus.

There are 175 galaxies in the SW01 sample that have measured emission line ratios with $S/N > 3 \sigma$. 
The SW01 sample covers substantially IR lower luminosities than the BGS or 1 Jy ULIRG samples and contains a 
large fraction of star-forming galaxies. In the 2-of-3 scheme,  the SW01 sample contains 78 (41.7\%) star-forming galaxies, 57 (30.5\%) starburst-AGN composites, 40 (21.4\%) Seyfert{\ts}2,  10 (5.3\%) Seyfert{\ts}1, and 2 (1.1\%) LINERs.  These
statistics include 12 double nuclei galaxies with consistent classifications for both nuclei (7/12 star-forming galaxies, 
3/12 composites, and 2/12 Seyfert 2 galaxies).   We do not include  4  double nuclei galaxies that have different spectral types for each nucleus.  Our results remain unchanged if we randomly assign these 4 galaxies the class of either nucleus.

Our classifications for the 1 Jy ULIRG, BGS, and SW01 samples are listed in Tables~\ref{tb1},~\ref{tb2} and \ref{tb3}, 
respectively.  For comparison, in Table~\ref{tb1} and \ref{tb2}, we also list the classifications given in \citet{veilleux99} 
using the traditional VO87 method.

\subsection{AGN Contribution}\label{dagn}
Ke06 showed that the SDSS galaxies form a mixing sequence between pure star-forming galaxies and pure AGN.  
They defined an empirical linear distance (${\rm D}_{SF}$) from the star-forming sequence for both the Seyfert and 
LINER branches on the \OIIIHb\ vs \OIHa\ diagnostic diagram.   The \OIIIHb\ vs \OIHa\ diagram was used to derive 
this distance because in this diagram, the Seyfert and LINER branches are clearly separated (unlike the \OIIIHb\ vs \NIIHa\ diagram 
where the Seyfert and LINER branches coincide).  Fig.~\ref{fig:sample} shows that unlike the optically-selected SDSS galaxies, 
the majority of IR-selected galaxies do not lie along the pure star-forming galaxy sequence; most IR galaxies lie in 
the composite and AGN regions of the diagnostic diagrams.  Because of this difference, it is more intuitive to think of the linear 
distance between the star-forming sequence and the AGN region on the \OIIIHb\ vs \OIHa\ diagnostic diagram as the 
relative contribution of an (${\rm D}_{AGN}$) for IR-selected galaxies.  
 
The quantity $D_{\rm AGN}$ can be defined other standard diagnostic diagrams with negligible difference.  In Appendix~\ref{defdagn}, we investigate alternative definitions of $D_{\rm AGN}$, and the relationship between $D_{\rm AGN}$ and spectral class.   Because $D_{\rm AGN}$ is a relative measure, our results remain the same regardless of how $D_{\rm AGN}$ is defined.   Note that because $D_{\rm AGN}$ is 
defined as a distance (in dex) in log line-ratio space, it does not give the fraction or a percentage of star-formation or AGN emission 
in a galaxy.  $D_{\rm AGN}$ gives a {\it relative} indication of the relative contribution of AGN to the EUV radiation field from galaxy 
to galaxy.  The absolute value of $D_{\rm AGN}$ is abstract, and $D_{\rm AGN}$ should be used only in relation to other galaxies. 
For example, a galaxy with $D_{\rm AGN}=0.6$ does not correspond to an AGN contribution of 60\%.  $D_{\rm AGN}$ is useful only for 
relative comparisons, for example,  a galaxy with $D_{\rm AGN}=0.6$ is likely to have a larger contribution from an AGN than a galaxy with a smaller $D_{\rm AGN}$.

\begin{figure}[!ht]
\epsscale{1.6}
\plotone{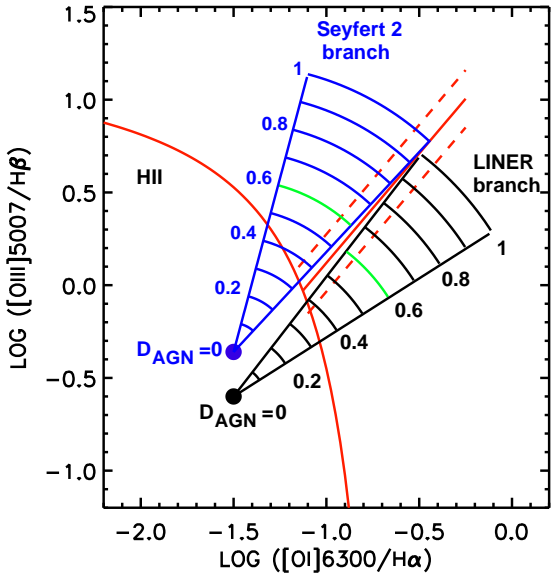}
\caption{\oi/H$\alpha$  versus \oiii/H$\beta$ diagnostic diagram showing the Ke06 classification 
scheme (red lines), and the distance to the peak of the AGN branch, $D_{\rm AGN}$.  Blue and black curves 
give lines of constant $D_{\rm AGN}$, respectively.  The green curves indicate the maximum $D_{\rm AGN}$ for 
starburst-AGN composite galaxies (defined in the  \nii/H$\alpha$ versus \oiii/H$\beta$ diagram).   Starburst 
galaxies and some starburst-AGN composite galaxies lie below the Ke06 maximum starburst line (red curve) in this diagram. }
\label{fig:DAGN}
\end{figure}

In Fig.~\ref{fig:DAGN}, we show $D_{\rm AGN}$ on the  \OIIIHb\ vs \OIHa\ diagram, relative to the Ke06 classification scheme (red lines).
By definition, $D_{\rm AGN}=D_{\rm SF}$ from Ke06.  Pure star-forming galaxies have $D_{\rm AGN}=0$.  The mixing sequence 
from pure star-forming galaxies to the tip of the AGN branch begins at $D_{\rm AGN} \geq 0$, and lies below the 
maximum starburst line (red curve in Fig.~\ref{fig:DAGN}).  Galaxies that are classed as  composites in the  \OIIIHb\ vs \NIIHa\ diagram 
have $D_{\rm AGN} \leq 0.5$ or 0.6 (green curve).  Galaxies that have $D_{\rm AGN}=1$ are likely to have line ratios that are strongly
dominated by an AGN, although some contribution from star formation can not be ruled out.  

\subsection{Merger Progress Tracers}\label{morphology}
We use two tracers of merger progress: merger morphology, and projected nuclear separation ({\it ns}).
We adopt the morphological classification scheme outlined in \citet{veilleux02}.  \citet{veilleux02} relate galaxy 
morphology to merger stage using numerical simulations of galaxy mergers \citep{barnes96}:

1. Wide binary: ``binary systems" with projected separation $ns >10$ kpc.

2. Close binary: ``binary systems" with projected separation $ns <10$ kpc.

3. Diffuse merger: single systems ($ns\sim 0$) with tidal features, and  with ${L_{K}}_{\rm 4kpc}/{L_{K}} < 1/3$, 
where ${L_{K}}_{\rm 4kpc}$ means the $K$-band luminosity within 4 kpc and ${L_{K}}$ is the total luminosity. 

4. Compact merger: single systems ($ns\sim 0$) with tidal features, and with ${L_{K}}_{\rm 4kpc}/{L_{K}} > 1/3$.

5. Old merger: single systems ($ns\sim 0$) with no unmistakable signs of tidal tails, yet have disturbed central
morphologies. 

We use the the projected separation measured in the $K_{\rm s}$-band and the length of the tidal tails measured 
in $R$-band by \citet{veilleux02} for the 1{\ts}Jy sample.   

For the BGS and SW01 samples, we use 2MASS $K_{\rm s}$-band images and the IRAF ``imcntr" task to measure 
the projected nuclear separations.  We use $R$-band DSS images to determine the morphological classification 
according to the Veilleux et al. scheme.  We search for companions within a 100 kpc radius.  The maximum angular 
radius available for 2MASS images is 10 arcminutes, giving galaxies at redshifts $z<0.0087$ images less than 
100 kpc wide.  For these low redshift galaxies, we use DSS images that have larger angular radii so that the 
100 kpc search radius criterion is always satisfied.   ``Isolated systems" defined in this way are therefore confined 
within the 100 kpc region.  It is possible that these systems are interacting with objects wider than 100 kpc.
For the few galaxies without available 2MASS images, we used the data from other images in NED or from the literature. 
These objects are noted in Tables~\ref{tb1},~\ref{tb2}, and \ref{tb3}.

For BGS LIRGs,  we find that 42/77 are mergers, and 35/77 are isolated systems or have companions outside the 
100 kpc search radius.  For the remaining 37 BGS objects, the image quality is too poor for morphological classification.  
We do not include these 37 galaxies in the morphological study.  Therefore the morphological classification for the  
BGS sample is incomplete.  We emphasize here that the main role of  the BGS sample in this study is to 
  supplement the 1{\ts}Jy ULIRG sample with lower luminosity objects, and to help create a larger non-ULIRGs 
  sample in \S~\ref{dagn_ns}.

The SW01 sample is not dominated by late stage 
mergers (class 3, 4, 5 above) and only 33/285 can be definitively identified as mergers with $ns\sim0$.  A significant 
fraction (156/285) of SW01 galaxies show no obvious signs of merging or interaction.  For statistical significance,  
we combine the ``diffuse", ``compact", and ``old" merger stages as one ``single merger" stage so that each class 
contains at least 10 galaxies. 
 
The final morphological classes for the SW01 and BGS samples are:  
 
1. Wide binary: ``binary systems" with projected separation $ns >10$ kpc.

2. Close binary: ``binary systems" with projected separation $ns <10$ kpc.

3. Single merger:  single systems ($ns\sim 0$) with tidal features or distorted nuclei recognizable in available images. 

4. Isolated system: systems that have no obvious signs of merging or interaction, within the 100 kpc image search radius.

There may be some overlap in the ``isolated" and ``single-merger" groups, as it is impossible to distinguish 
whether a galaxy is truly ``isolated" or is simply at the end of the merging stage where all the tidal features 
disappear (or are too faint to be observed).  For example, there are 7 single nucleus ULIRGs in the SW01 sample
that do not show obvious signs of merging in available NED images.  However, 1{\ts}Jy  $R$-band images show 
remnant signs of tidal activity for 3 of the single nuclei galaxies.  

Besides the uncertainty in distinguishing the ``isolated systems" from ``single mergers", there 
are some classical drawbacks in using projected separation as a tracer for merger processes:

a) Projection effects may randomize the results.  However, for a large sample, projection effects should be 
statistically unimportant; on average, interacting systems with {\it ns} $\lesssim $ few kpc are likely to be at a later 
stage in the merger process than systems with {\it ns} $\gtrsim $  a few 10~kpc.

b) Merger models such as \citet{barnes96} show that {\it ns} is not necessarily a linear function of time: the projected 
separation decreases in a period of close contact (``first-pass") and then increases again before the nuclei finally merge. 

c) Projected nuclear separation can not trace merger progress in multiple mergers 
of more than two galaxies unless the time between current mergers and that of the 
former mergers is sufficiently large to enable any merger-induced star formation 
and AGN activity to subside \citep{borne00}. This issue may be a potential problem 
for the SW01 sample, in which some galaxies are in the Hickson Compact Groups
\citep{garcia93}.  In these groups, the IR emission may be triggered by the weak interaction between
 the group members.  For the 1{\ts}Jy sample, multiple mergers are not a major 
 concern because deep images indicate that only 5/118 ($<5$\%) are {\it possible} multiple mergers \citep{veilleux02}.

With these caveats in mind, we calculate the projected separation between paired 
galaxies in our samples using 2MASS $K_{\rm s}$-band images for the BGS and SW01 samples, 
and the \citet{veilleux02} $K_{\rm s}$-band projected separation measurements for the 1{\ts}Jy sample.

\section{RESULTS}\label{results}

\subsection{Spectral Type as a Function of IR luminosity}\label{speclir}

Our samples span different luminosity ranges and have different IR color selection criteria.   
To examine how spectral type changes with IR luminosity, we combine the 1{\ts}Jy ULIRG sample 
with the  BGS sample to cover a broad IR luminosity range and to facilitate comparisons with the classical 
\citet{veilleux99} result.   We consider the SW01 sample separately in this case because the warm color 
criterion of the SW01 sample may affect how the spectral type changes with IR luminosity.   

\begin{figure}[!ht]
\epsscale{1}
\plotone{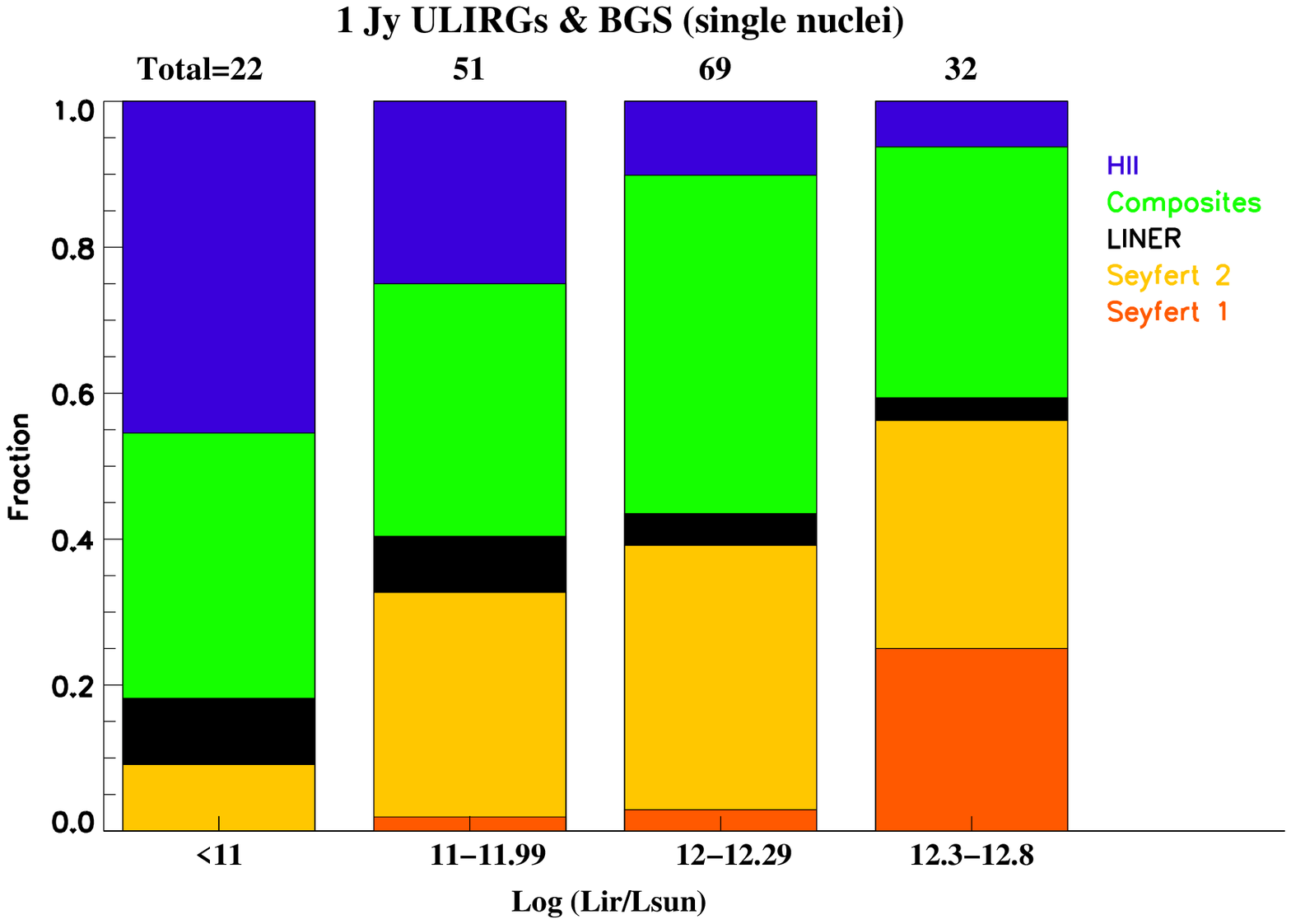}
\caption{Spectral type (2-of-3 criterion) as a function of  $L_{\rm IR}$ for the   
1{\ts}Jy ULIRG sample \citep{veilleux99} and the LIRGs in the BGS sample \citep{veilleux95}. 
Similar to \citet{veilleux99}, double nucleus objects are excluded in this figure.  The number of galaxies 
contained in each bin is marked on top of the histogram. The luminosity bins are labeled at the bottom. 
Throughout this paper we will always use grey for ambiguous, blue for HII-region galaxies, green for 
composites, black for LINERs, yellow for Seyfert{\ts}2 galaxies and orange for Seyfert{\ts}1 galaxies, 
unless otherwise specified. The result of applying  the stringent 3-of-3 criterion is presented in Appendix~\ref{ambissue}.}
\label{fig:lirub1}
\end{figure}

Fig.~\ref{fig:lirub1} shows the optical spectral type as a function of $L_{\rm IR}$ for the combined 
1{\ts}Jy and BGS samples. The results are also tabulated in Table~\ref{tb4}.  It is obvious from this figure that LINERs are rare compared with other spectral classes
 in our IR-luminous samples, with 
only 3\% LINERs in the 1{\ts}Jy sample and 7\% in the BGS sample.  The 1{\ts}Jy ULIRGs sample may lack LINERs because 
 this sample is selected at larger redshifts than the SDSS.  Ke06 showed that the fraction of LINERs in the SDSS falls 
 at $z>0.1$ due to incompleteness.  The lack of LINERs in SW01 may be at least partly caused by the warm selection criterion, as the the ``warm"  criterion selects against LINERs \citep{kewley01a}.

We find few IR-luminous LINERs found in our samples. A majority (72/79) 
of the objects previously classified as LINERs are 
classified as composite galaxies and Seyfert{\ts}2s using the Ke06 SDSS-based classification scheme.  LINERs in the SDSS-based classification scheme contain an older stellar population and an AGN with a lower accretion rate than Seyfert{\ts}2 galaxies.  The lack of such LINERs in IR samples indicates that (a) bona fide IR-luminous LINERs are quite rare --- only 3/108 in the 1{\ts}Jy ULIRG sample and 5/97 in the BGS samples, and (b) in most cases, LINERs do not contain the intense star formation and/or the strong dust-reprocessed AGN emission responsible for the IR luminosities seen in U/LIRGs.  
We discuss the rare IR-selected LINERs in more detail in \S~\ref{liners}.

\begin{figure}[!ht]
\epsscale{1}
\plotone{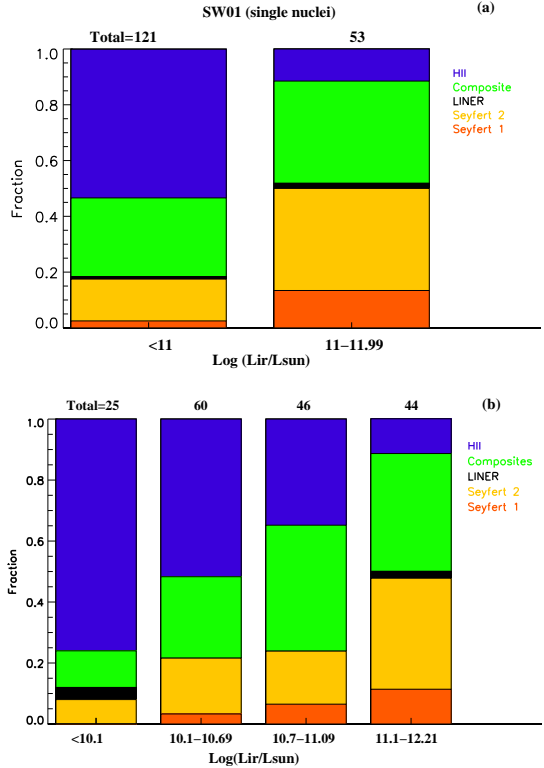}
\caption{  
Optical classification results as a function of $L_{\rm IR}$ for the SW01 sample.  The SW01 sample is  dominated by moderate IR luminosity $L_{\rm IR}<10^{11}L_{\odot}$  sources, with $\sim$25\% LIRGs, 
and very few ULIRGs ($<$2\% = 3 objects).  For comparison with Figure~\ref{fig:lirub1}, Panel (a) divides the SW01 sample
into the same two lower luminosity bins (only 3 SW01 objects have $L_{\rm IR}>10^{12}L_{\odot}$ in the bin 12$<$ log($L_{\rm IR}/L_{\odot}$)$<$12.21).   Panel (b) provides a finer grouping into four $L_{\rm IR}$ bins to gain more resolution.  
The sub-bins are chosen in such a way that each bin contains at least 20 galaxies. }
\label{fig:swlir}
\end{figure}

Interestingly, we find that Seyfert{\ts}1 galaxies favor higher $L_{\rm IR}$, consistent with previous studies suggesting 
that Seyfert{\ts}2 galaxies have weaker mid-IR luminosities than Seyfert{\ts}1 galaxies 
\citep{heckman95,maiolino95,giuricin95}  \citep[however, see][]{bonatto97,haas07}.   However, because the 1{\ts}Jy ULIRG + BGS samples only contain a total of 11 Seyfert{\ts}1 galaxies, a larger number of IR-luminous Seyfert{\ts}1s is required to 
determine the significance of this result.

The spectral type as a function of $L_{\rm IR}$ for the SW01 sample is shown in Fig.~\ref{fig:swlir} and Table~\ref{tb5}.   Because the SW01 sample
covers a lower IR luminosity range, the SW01 sample is intended to extend the 1 Jy ULIRG and BGS samples to lower luminosities, rather than to serve as a comparison sample.  Nevertheless, in panel (a) of Fig.~\ref{fig:swlir},  the  $L_{\rm IR}$ bins are chosen to be the same as the first two bins of Fig.~\ref{fig:lirub1} to facilitate 
comparisons between the warm (SW01) and non-warm (BGS and 1 Jy ULIRG) samples over the $L_{\rm IR}$ range where all
samples are well-defined.

Since our samples contain a large number of objects with $L_{\rm IR}/L_{\odot}< 10^{10}$, $L_{\rm IR}$ is divided into sub-bins in panel (b) to improve resolution.  Bin sizes are chosen to to ensure that each bin contains at least 20 galaxies.   In the SW01 sample, the AGN fraction increases and the fraction of starbursts decreases as $L_{\rm IR}$ becomes larger, 
similar to the trend in the 1{\ts}Jy ULIRG +  BGS samples.  As shown in panel (b) of Fig.~\ref{fig:swlir}, the lowest luminosity bin  
($L_{\rm IR}/L_{\odot}< 10^{10}$) has the largest fraction of star-forming galaxies.

\subsection{$D_{\rm AGN}$ as a function of $L_{\rm IR}$}\label{dagn_lir}
\begin{figure}[!ht]
\epsscale{1.23}
\plotone{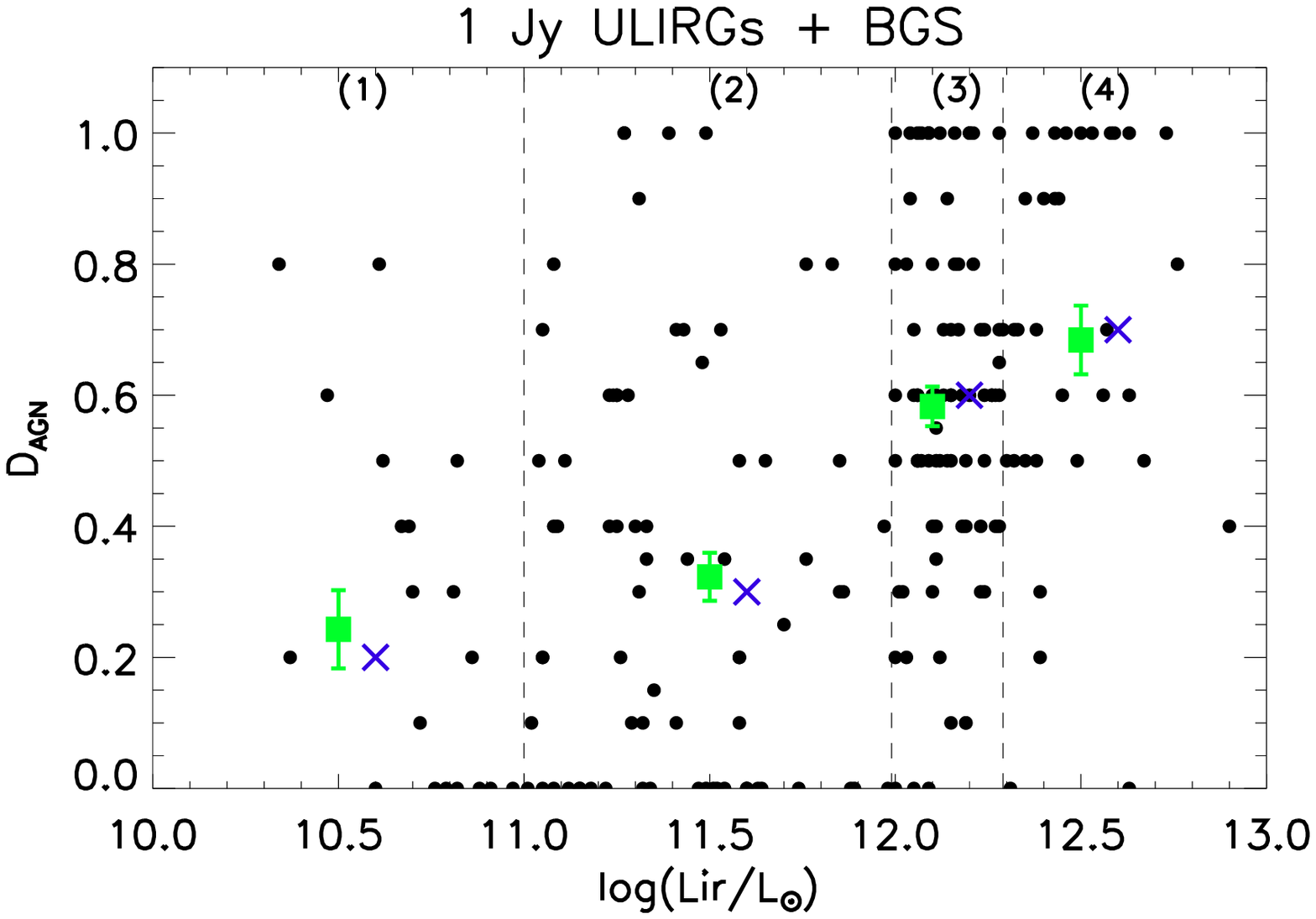}
\caption{ $D_{\rm AGN}$ as a function of $L_{\rm IR}$ for the 1{\ts}Jy ULIRG + BGS sample.  The mean, median, and standard error
of the mean for the same 4 IR luminosity bins used in Fig.~\ref{fig:lirub1} are plotted in green. 
Blue crosses give the median values.  The probability that the $D_{\rm AGN}$ distribution in consecutive bin pairs is drawn from the same parent population is given in \S~\ref{dagn_lir}. The rise in $D_{\rm AGN}$ from luminosity bin (2) to (3) and 
bin (3) to (4) is significant at the 99\% and 90\% confidence levels, respectively}
\label{fig:dlirub}
\end{figure}

\begin{figure}[!ht]
\epsscale{1.23}
\plotone{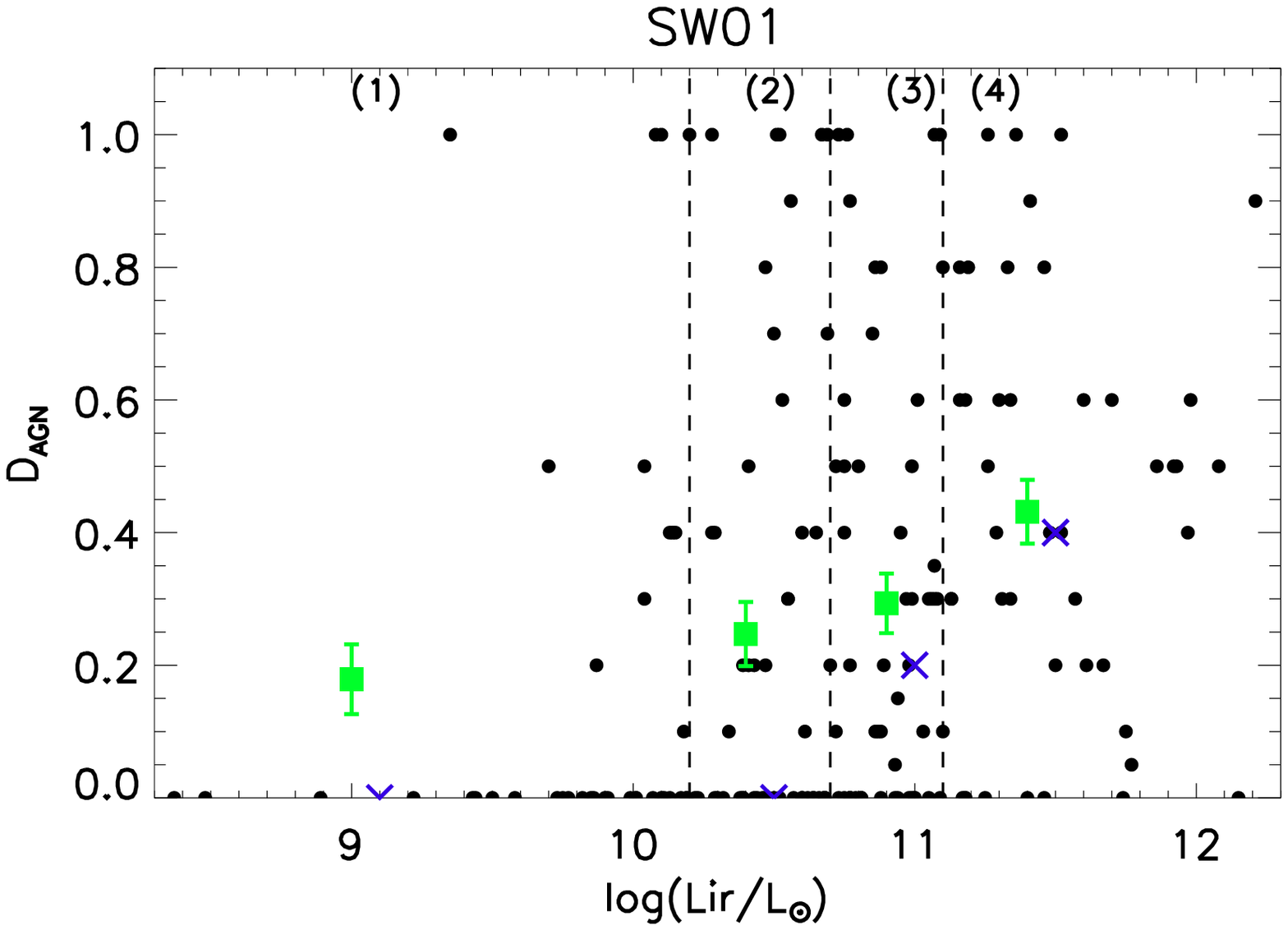}
\caption{$D_{\rm AGN}$ as a function of $L_{\rm IR}$ for the SW01 sample.   The mean, median,  and standard error
of the mean for the same 4 IR luminosity bins used in Fig.~\ref{fig:swlir} panel (b) are plotted in green. Blue crosses are the median values.  The probability that the $D_{\rm AGN}$ distribution in consecutive bin pairs is drawn from the same parent population is given in \S~\ref{dagn_lir}. The rise in $D_{\rm AGN}$ from luminosity bin (3) to (4) is significant at the 98\% confidence level}
\label{fig:dlirsw}
\end{figure}

Spectral types only reflect the galaxy's general position on the BPT diagrams.  To investigate the relative contribution 
from an AGN as a function of IR luminosity,  we show in Fig.~\ref{fig:dlirub}  and  Fig.~\ref{fig:dlirsw} $D_{\rm AGN}$ as a 
function of $L_{\rm IR}$ for the combined 1{\ts}Jy ULIRG + BGS sample, and the SW01 sample, respectively. 
The mean, median, and standard errors for $D_{\rm AGN}$ are calculated for the same four luminosity ranges
as used in Fig.~\ref{fig:lirub1} and Fig.~\ref{fig:swlir}b,  for the combined 1{\ts}Jy ULIRG + BGS sample and the SW01 sample, 
respectively.  
To test whether the distribution of $D_{\rm AGN}$ in each $L_{\rm IR}$ bin can be drawn randomly from the 
same parent population, we perform the two-sample Kolmogorov-Smirnov (KS) test to each pair of neighboring $L_{\rm IR}$ bins. 
For the $L_{\rm IR}$ bin pairs (1-2), (2-3), (3-4) in Fig.~\ref{fig:dlirub}, 
the KS test gives a probability $P_{\rm null}$ = 0.08, 3.98E-6, and 0.16 that the distributions are drawn from the 
same parent population, indicating a statistically significant change in the $D_{\rm AGN}$ distribution at the 90\% confidence level 
between $10^{10}<L_{\rm IR}/L_{\odot}<10^{12.3}$ for the 1{\ts}Jy ULIRG + BGS sample.

For the SW01 $L_{\rm IR}$ bin pairs (1-2), (2-3), (3-4) in Fig.~\ref{fig:dlirsw},  
the KS test gives a probability $P_{\rm null}$ = 0.48, 0.28, and 0.02 that the distributions are drawn from the same 
parent population, indicating a statistically significant change in the $D_{\rm AGN}$ distribution at the 98\% confidence level 
between $10^{10.7}<L_{\rm IR}/L_{\odot}<10^{12.3}$.  The rise in $D_{\rm AGN}$ at lower luminosities is not statistically
significant for the SW01 sample.

We conclude that the rise of  $D_{\rm AGN}$ with $L_{\rm IR}$ is significant for all three samples at  $L_{\rm IR}/L_{\odot}> 10^{11.0}$.   Previous studies indicate that the fraction of AGN increases at large $L_{\rm IR}$ \citep[e.g.,][]{armus90,veilleux95,goto05}.  Figures~\ref{fig:dlirub} and \ref{fig:dlirsw} confirm these results.


\subsection{Spectral type as a function of morphology/merger stage\label{spec_vs_morph}}

\begin{figure}[!ht]
\epsscale{1.1}
\plotone{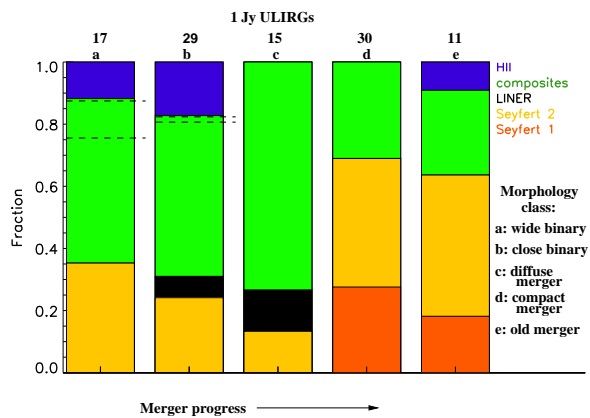}
\caption{Optical spectroscopic classification as a function of merger morphology for the 1{\ts}Jy ULIRG sample.  The merger 
progresses from left to right: a. wide binary,\  b. close binary,\  c. diffuse merger,\ d. compact merger,\ e. old merger.  
No "isolated" class is defined for the 1{\ts}Jy morphology class since there is only one isolated object in this sample.
The number of galaxies in each morphology class is marked on top of each bin, and is given in Table~\ref{tb6}. The 
fraction of composite galaxies reaches a peak at the diffuse merger stage.  The fraction of AGN-dominated 
(Seyfert{\ts}2 and Seyfert{\ts}1) rises at later merger stages (compact and old merger). 
 Dashed lines give the range of uncertainty in the fraction of HII and composites associated with the double nuclei 
galaxies with different spectral types for each nucleus (see Appendix~\ref{dnissue} . } 
\label{fig:moru}
\end{figure}

Following the morphological definitions in \S~\ref{morphology}, in Fig.~\ref{fig:moru} we give
 the spectral type as a function of merging stage for the 1{\ts}Jy ULIRG sample.     
 The dashed line indicates the uncertainty in classification associated with the double nuclei galaxies that have 
 differing classifications for each nucleus.   These galaxies are discussed in more detail in Appendix~\ref{dnissue}.

\begin{figure}[!ht]
\epsscale{1.1}
\plotone{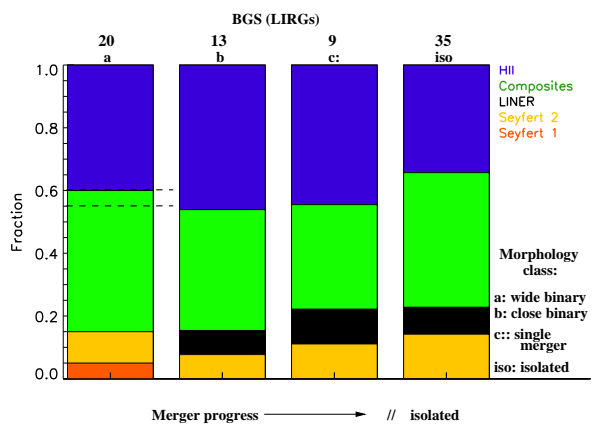}
\caption{Spectral type as a function of morphology for the LIRGs in the BGS sample. 
The merger progresses from left to right:\ a. wide binary,\ b. close binary,\ c:. single-merger, iso. isolated system.
 Dashed lines give the range of uncertainty in the fraction of HII and composites caused by the double nuclei 
galaxies with different spectral types for each nucleus.   The single-merger stage combines the diffuse, compact, and old merger stages to ensure sufficient numbers of 
objects for statistically meaningful comparisons between the morphological types.  There may be overlap 
between single merger and isolated stages due to surface brightness effects.
}\label{fig:morbgs}
\end{figure}
\begin{figure}[!ht]
\epsscale{1.1}
\plotone{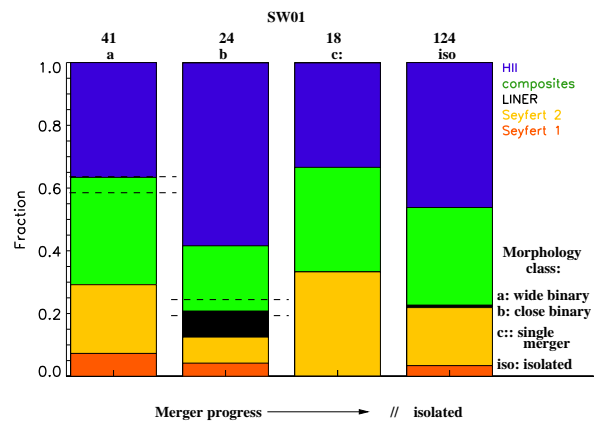}
\caption{Optical spectroscopic classification as a function of merger morphology for the SW01 sample.  
The merger progresses from left to right:\ a. wide binary,\ b. close binary,\  c:. single merger,\ d. isolated.   
Dashed lines give the range of uncertainty in the fraction of HII and composites (in stage a), 
 composite and Seyferts (in stage b) caused by the double nuclei 
galaxies with different spectral types for each nucleus.
The single-merger stage combines the diffuse, compact, and old merger stages to ensure sufficient numbers of 
objects for statistically meaningful comparisons between the morphological types.  The SW01 sample contains a 
substantial fraction (156/285$\sim 54$\%) of apparently isolated galaxies.  There may be overlap between single 
merger and isolated stages due to surface brightness effects.  The fraction of composites is not significant as 
compared to Fig.~\ref{fig:moru}.}\label{fig:morsw}
\end{figure}
\begin{figure}[ht]
\epsscale{1.1}
\plotone{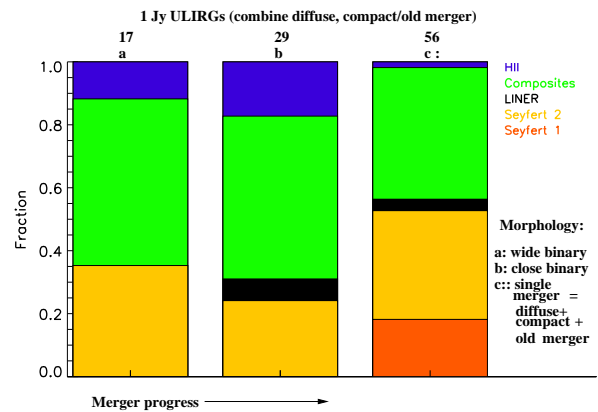}
\caption{ Optical spectroscopic classification as a function of merger morphology for the 1{\ts}Jy ULIRG sample, after 
combining the diffuse, compact and old mergers as one single merger category.}
\label{fig:moru2}
\end{figure}

The key significant features in Figure~\ref{fig:moru} are:

1)  The fraction of starburst-AGN composite galaxies reaches a peak at the diffuse merger stage, indicating that both merger-induced starburst activity, and AGN fueling contribute to the total energy budget at the diffuse merger stage.  

2) The fraction of starburst-AGN composite galaxies falls sharply at the later merger stages (compact and old merger), giving 
rise to a larger portion of Seyfert galaxies rises in these later merger stages.  We suggest that starburst-AGN composite galaxies 
evolve into Seyferts as merger-induced starburst activity subsides.
 
Figure~\ref{fig:moru} also shows potentially interesting trends that are limited by small numbers for the 1 Jy ULIRG sample:

(1) The fraction of Seyfert galaxies appears to decrease ($41\pm16$\% c.f. $13\pm 9$\%) between the wide binary stage (stage a) to the diffuse merger stage (stage c).

(2) Seyfert 1 galaxies only occur at the later merger stages (d and e in Figure~\ref{fig:moru} )   
 
A larger sample of ULIRGs is required to verify these two trends.

We show the spectral type as a function of merger morphology for the BGS and SW01 samples in Figures~\ref{fig:morbgs} and
\ref{fig:morsw} respectively. The results are listed in Table~\ref{tb7} and  Table~\ref{tb8}. Clearly, for objects with $L_{\rm IR}$ lower than that of ULIRGs, there is no strong change in spectral classification as a function of merger progress.   However, the BGS and SW01 samples do not contain a sufficient number of galaxies in the single merger stage that have deep enough images to allow separation into diffuse, compact, and old 
merger stages where the largest changes in spectral class occur for ULIRGs.  The fraction of merging galaxies in the SW01 and BGS samples is 45\% and 54\%, respectively, compared with $99$\% for the 1{\ts}Jy ULIRG sample.  In addition, most of the merging systems in the SW01 and BGS samples are in the earlier, binary stages rather than the advanced single-merger stages in the 1{\ts}Jy  ULIRG sample (11\% in the SW01 sample, and 12\% in the BGS sample versus 50\% in the 1{\ts}Jy ULIRG sample).   
If we combine objects in the 1{\ts}Jy ULIRG sample which have diffuse, compact, and old merger types into one single 
merger class (Fig.~\ref{fig:moru2}), the changes seen in Fig.~\ref{fig:morsw} disappear.  These results highlight the 
need for sensitive $K-$ and $B-$band imaging to distinguish between the three late merger stages (diffuse, compact, and old merger) in non-ULIRG samples.

To conclude, the 1{\ts}Jy ULIRG sample shows a marked change in galaxy spectral type as a function of 
merger progress, especially, composite galaxies dominate the diffuse merger stage.  It is unclear whether such a change occurs in the lower luminosity BGS and SW01 samples.  We note that the relative lack of galaxies at late merger stages in the lower luminosity samples may indicate that a ULIRG phase occurs before or at the single nucleus stage, in at least some IR-selected galaxies.

\subsection{$D_{\rm AGN}$ and $L_{\rm IR}$ versus $ns$}\label{dagn_ns}

\begin{figure}[!ht]
\epsscale{1.1}
\plotone{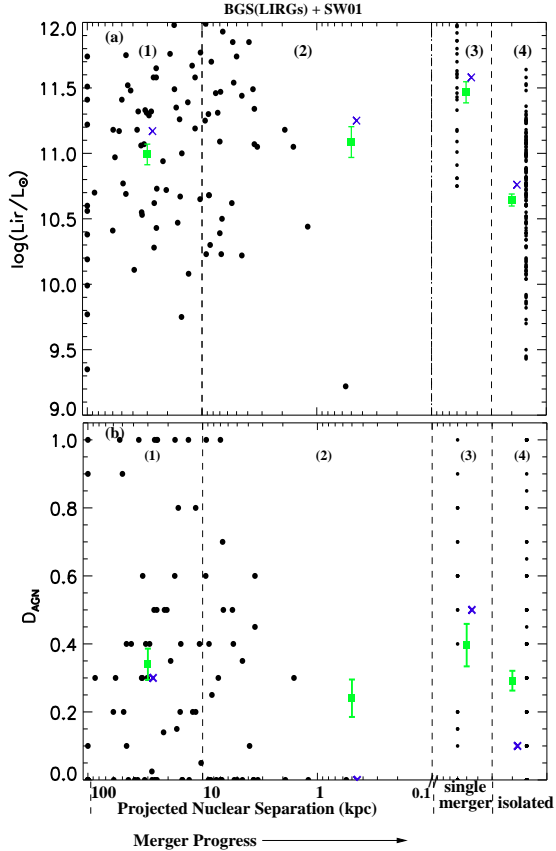}
\caption{
(a) $L_{\rm IR}$ as a function of projected nuclear separation for the BGS (LIRGs only) and SW01 samples combined. 
(b) Distance from the starburst sequence $D_{\rm AGN}$ as a function of projected nuclear separation 
for the combined BGS (LIRGs only) and SW01 samples.   The mean and 1$\sigma$ standard deviation
of the mean for $D_{\rm AGN}$ are plotted as green squares.  Median values are shown as blue crosses. 
The vertical regions from left to right are:  
(1) binary systems with nuclear separation  $ns > 10${\ts}kpc. 
(2) binary systems with nuclear separation $ns < 10${\ts}kpc. 
(3) single-merger systems. 
(4) isolated systems. 
The BGS and SW01 samples are combined in order to get a larger non-ULIRG sample.
For the 21 overlap objects in the two samples, mean values of $D_{\rm AGN}$ are taken. 
KS tests for the significance of the difference in y-axis distributions are given in \S~\ref{dagn_ns}.
 }
\label{fig:lirdagn_bgsw}
\end{figure}

\begin{figure}[!ht]
\epsscale{1.2}
\plotone{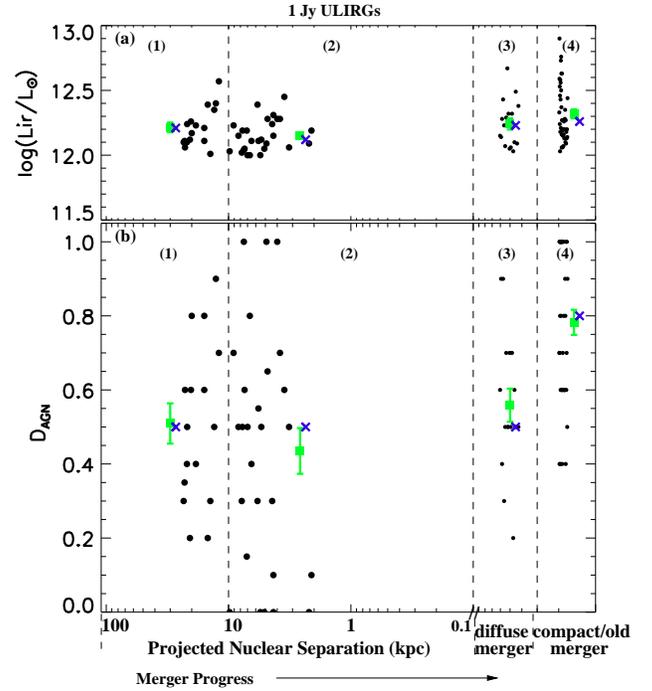}
\caption{   
(a) $L_{\rm IR}$ as a function of projected nuclear separation for the 1Jy ULIRGs. 
(b) Distance from the starburst sequence $D_{\rm AGN}$ as a function of projected nuclear separation for the 1{\ts}Jy ULIRG sample.  The means and  1$\sigma$ standard deviation of the mean for $D_{\rm AGN}$ are plotted as green squares. Median values are shown as blue crosses.  
The vertical regions from left to right are:  
(1) binary systems with nuclear separation  $ns > 10${\ts}kpc. 
(2) binary systems with nuclear separation $ns < 10${\ts}kpc. 
(3) single-mergers with diffuse nuclei.
(4) single-mergers  with compact/old nuclei. 
KS tests for the significance of the difference in y-axis distributions are given in \S~\ref{dagn_ns}.
}
\label{fig:lirdagn_ulig}
\end{figure}


To investigate the merger progress in a morphology-independent way, we compare the {\it relative} contribution of an AGN ($D_{\rm AGN}$) and $L_{\rm IR}$ as a function of projected nuclear separation ($ns$).   We first investigate the non-ULIRG samples by combining the BGS and SW01 samples to (a) reach statistically significant conclusions (thus minimizing projection effects) and (b) span a broad range of $L_{\rm IR}$.  The $ns$ values for both samples are measured using 2MASS or DSS images, and the two samples exhibit similar behavior in spectral type versus morphology as discussed in \S~\ref{spec_vs_morph}. 

In Figure~\ref{fig:lirdagn_bgsw}(a), we show the infrared luminosity, $L_{\rm IR}$, of the merger pair as a function of nuclear separation. Fig.~\ref{fig:lirdagn_bgsw}(b) gives $D_{\rm AGN}$ as a function of nuclear separation for the BGS (LIRGs) and the SW01 samples combined.   These figures indicate that $D_{\rm AGN}$ is constant at $\sim 0.3$ within the errors at all merger stages in these 2 samples.  The mean value of $L_{\rm IR}$ is constant within the errors through the wide and close pair stages, and then rises by a factor of $\sim 2$ in the single merger stage.  We performed KS tests to determine the significance of the  distribution of $D_{\rm AGN}$ and $L_{\rm IR}$ as a function of projected separation.  For the distribution of data within adjacent $L_{\rm IR}$ bins (1) and (2), (2) and (3), and (3) and (4) in Fig.~\ref{fig:lirdagn_bgsw}a, the KS test indicates a probability $P_{\rm null}$ = 0.89, 0.04, and 3.8E-9 respectively that the adjacent data sets are drawn from the same parent population.
For the adjacent $D_{\rm AGN}$ bins (1) and (2), (2) and (3), (3) and (4) in Fig.~\ref{fig:lirdagn_bgsw}b, 
the KS test indicates a probability $P_{\rm null}$ = 0.13, 0.18, and 0.16 that the adjacent data sets are drawn from the same
parent population.  We conclude that both the rise in $L_{\rm IR}$ from close pair to single merger stage 
and the fall in $L_{\rm IR}$ from single merger to isolated stages are significant at the 95\% and 99.99\% level respectively.
There are no statistically significant changes in $D_{\rm AGN}$ as a function of projected separation for the non-ULIRG BGS and SW01 samples.

In Figure~\ref{fig:lirdagn_ulig}(a) and Figure.~\ref{fig:lirdagn_ulig}(b) we show the $L_{\rm IR}$ and $D_{\rm AGN}$ versus projected separation for the 1{\ts}Jy ULIRG sample.  Like the lower luminosity pairs, $D_{\rm AGN}$ is constant within the errors as a function of projected separation until the diffuse merger stage.  The mean $D_{\rm AGN}$ is larger than seen in the lower luminosity BGS$+$SW01 sample ($\sim$0.52 c.f. 0.3).   The most obvious difference in the behavior of ULIRGs with projected separation compared with the lower luminosity pairs, occurs in the final compact/old merger phase where ULIRGs show a rather dramatic increase in $D_{\rm AGN}$ to a mean value of $\sim 0.8$.  This rise is statistically significant at the 95\% level; the KS test indicates a probability $P_{\rm null}$ = 0.46, 0.51, and 0.05 that adjacent data sets  (1-2), (2-3), (3-4) are drawn from the same parent population.   This rise in $D_{\rm AGN}$ is consistent with the rise in the Seyfert fraction with merger progress seen in \S~\ref{spec_vs_morph}, particularly in the emergence of Seyfert{\ts}1s at the compact and old merger stages.   The rise in $D_{\rm AGN}$ is not accompanied by a significant rise in the mean $L_{\rm IR}$ (the KS test indicates a probability $P_{\rm null}$ =  0.52, 0.12, and 0.84 that the adjacent data sets  (1-2), (2-3), (3-4) are drawn from the same parent population).   

Our results confirm previous observational \citep[e.g.,][]{sanders88a} and theoretical \citep[e.g.,][]{mihos94} studies showing that 
the maximum $L_{\rm IR}$ is produced close to the time when the two nuclei merge.

\section{DISCUSSION}\label{discussion}

\subsection{IR-selected LINERs}\label{liners}

We find that IR luminous LINERs are rare compared to other optical spectral types, 
even in the low-redshift BGS sample.  A majority of the previously classified IR-luminous LINERs 
are now reclassified as starburst-AGN composites in the new SDSS optical spectroscopic classification scheme.
This result prompts us to discuss the following issues: \ 
(1) The relationship between the new class of LINERs and starburst-AGN composite objects. \  
(2) The nature of IR-luminous LINERs compared with the LINERs from the SDSS. 
We discuss each of these issues separately below. 

(1)  \citet{kewley06} have shown that the host properties of 
composite galaxies are intermediate between those of AGN and high metallicity star-forming galaxies, and that 
LINERs form a unique, coherent class that is distinct from Seyferts, star-forming galaxies and composites.  
Specifically, LINERs are older, more massive, less dusty and less concentrated than Seyfert galaxies. 
These optical results highlight the intrinsically different nature of LINERs and composites.

\begin{figure}[!ht]
\epsscale{1.56}
\plotone{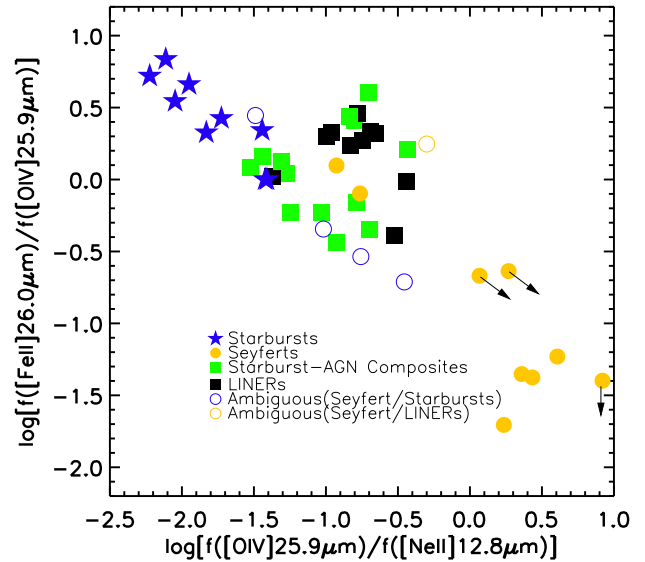}
\caption{Mid-infrared diagnostic from \citet{sturm06}, with objects re-classified using the Ke06 new scheme.
The data were provided by Sturm, E.  The symbol meanings are labeled on the plot.  The bracket after the ambiguous
class indicates between which classes the ambiguity lies. }
\label{fig:sturm}
\end{figure}

Here we investigate whether similar differences exist in popular mid-IR diagnostics.  In Figure~\ref{fig:sturm}, we investigate the positions of galaxies on the [FeIII]~$26.0 \mu m$/[OIV]~$25.9 \mu m$ versus [OIV]~$25.9 \mu m$/[NeIII]~$12.8 \mu m$ diagnostic diagram.  This diagram separates star-forming galaxies from those dominated by an AGN.   Previously, IR-luminous LINERs 
occupied a region in between starburst and Seyfert galaxies, whereas IR-faint LINERs occupied a region offset from the starburst-Seyfert sequence \citep{sturm06}.    In our new classification scheme,  only 1 out of the 16 IR-luminous LINERs are classified as LINERs (the rest 15
 objects are 8 composites, 1 Seyfert 2s, and 6 ambiguous classes between HII and Seyfert 2s), while 10 out of the 17 IR-faint LINERs remain LINERs (the rest 7 objects are ambiguous classes between LINERS and Seyfert 2s).    Figure~\ref{fig:sturm} indicates that the majority (9/13) of the composites lie in along 
a mixing sequence connecting star-forming galaxies and Seyfert{\it s}.  By contrast, most (8/10) of the LINERs lie offset from the mixing sequence. These results are consistent with our view of optically selected LINERs as a different class of objects from starburst-AGN composites.

(2) There is growing evidence that a large fraction of optically selected LINERs are low luminosity AGN 
with low accretion rates \citep[e.g.,][]{ho99, quataert01, barth02}.   The SDSS LINER population studied by \citet{kewley06} 
supports this AGN nature. They show that at fixed $L$\oiii/$\sigma^4$ (an indicator for the black hole accretion
 rate), all differences between Seyfert and LINER host properties disappear.  With this
 interpretation of LINERs, it is not surprising that we find so few IR-luminous LINERs because  
 ULIRGs tend to harbor AGN that favor high-accretion rates \citep{weedman83, filipenko85, sanders88a}.
Theory supports this interpretation; numerical simulations predict that the black hole accretion rate 
rises rapidly during the merger process when enormous quantities of gas flows into the central regions of the 
merging galaxies  \citep[e.g.,][]{tani99, hopkins05}. 

\begin{figure}[!ht]
\epsscale{1.4}
\plotone{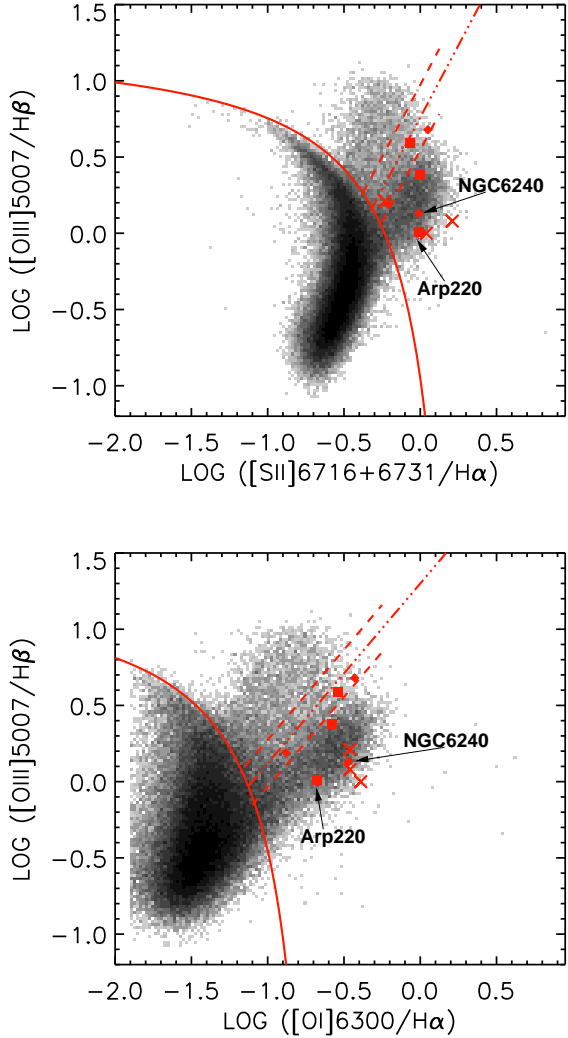}
\caption{   
The positions of the 9 LINERs on the  \sii/H$\alpha$ versus \oiii/H$\beta$ and \oi/H$\alpha$ versus \oiii/H$\beta$ diagrams over-plotted on the SDSS data from \citet{kewley06}.
The 3 IR-luminous LINERs in the 1{\ts}Jy ULIRG sample are shown as squares, the 3 IR-luminous LINERs in the 
SW01 sample are shown as diamonds, and the 3 crosses are the IR-luminous LINERs from the BGS sample.
}
\label{fig:linear1}
\end{figure}

What about the few IR-luminous LINERs that do exist in our samples?  Do they belong to the same class as 
optical LINERs?   In Table~\ref{tb9} we list all the possible LINERs found in our combined 1{\ts}Jy ULIRG + BGS + SW01 sample.   
Altogether there are 9 objects in the three samples that fall into our LINER class,  however, only 4/9 of these 
LINERs (Arp 220 in the 1{\ts}Jy ULIRG sample, NGC 6240 in the SW01/BGS samples and another 3 in the BGS sample) 
can be ``safely" classified as LINERs.  The other 5/9 LINERs lie near the 0.1{\ts}dex error region of the Seyfert-LINER boundaries,
and may be Seyfert objects or intermediate between Seyfert and LINER types.   This can be clearly seen in Figure~\ref{fig:linear1}, 
where all 9 LINERs are over-plotted with the SDSS data from \citet{kewley06} on the \sii/H$\alpha$ versus \oiii/H$\beta$ and  \oi/H$\alpha$ versus \oiii/H$\beta$ diagnostic diagrams.

In Figure~\ref{fig:linear1}, we indicate the positions of two well-studied IR-luminous LINERs: NGC 6240 and Arp 220.
 Recent high resolution X-ray 
 and radio data for NGC 6240 \citep{lira02, komossa03, iono07} and Arp 220 \citep{clements02, ptak03, downes07} 
provide convincing evidence for the existence of AGN in these 2 objects.   However, the LINER emission may be excited by 
other ionization processes.  Both of these two LINERs show evidence for starburst-driven superwinds and/or shocks that may
dominate the EUV emission of these galaxies \citep{heckmanet87, tecza00,lira02, mc03, iwa05}.  Further investigation into 
the EUV power source of IR-luminous LINERs is required to draw robust conclusions about the nature of the few IR-luminous LINERs in our samples.

To summarize, the rarity of IR-luminous LINERs is consistent with the picture that most LINERs are (a) excited 
by low Eddington rate black holes and (b) reside in galaxies with an aged stellar population.   The special cases of 
Arp 220 and NGC 6240 may indicate a contribution from different LINER ionization sources: either starburst 
superwinds or shocks driven by galaxy collisions.

\subsection{Are starburst-AGN composites playing the role of ``bridging" in LIRGs and ULIRGs? }

\begin{figure*}[!bthp]
\epsscale{1.14}
\plotone{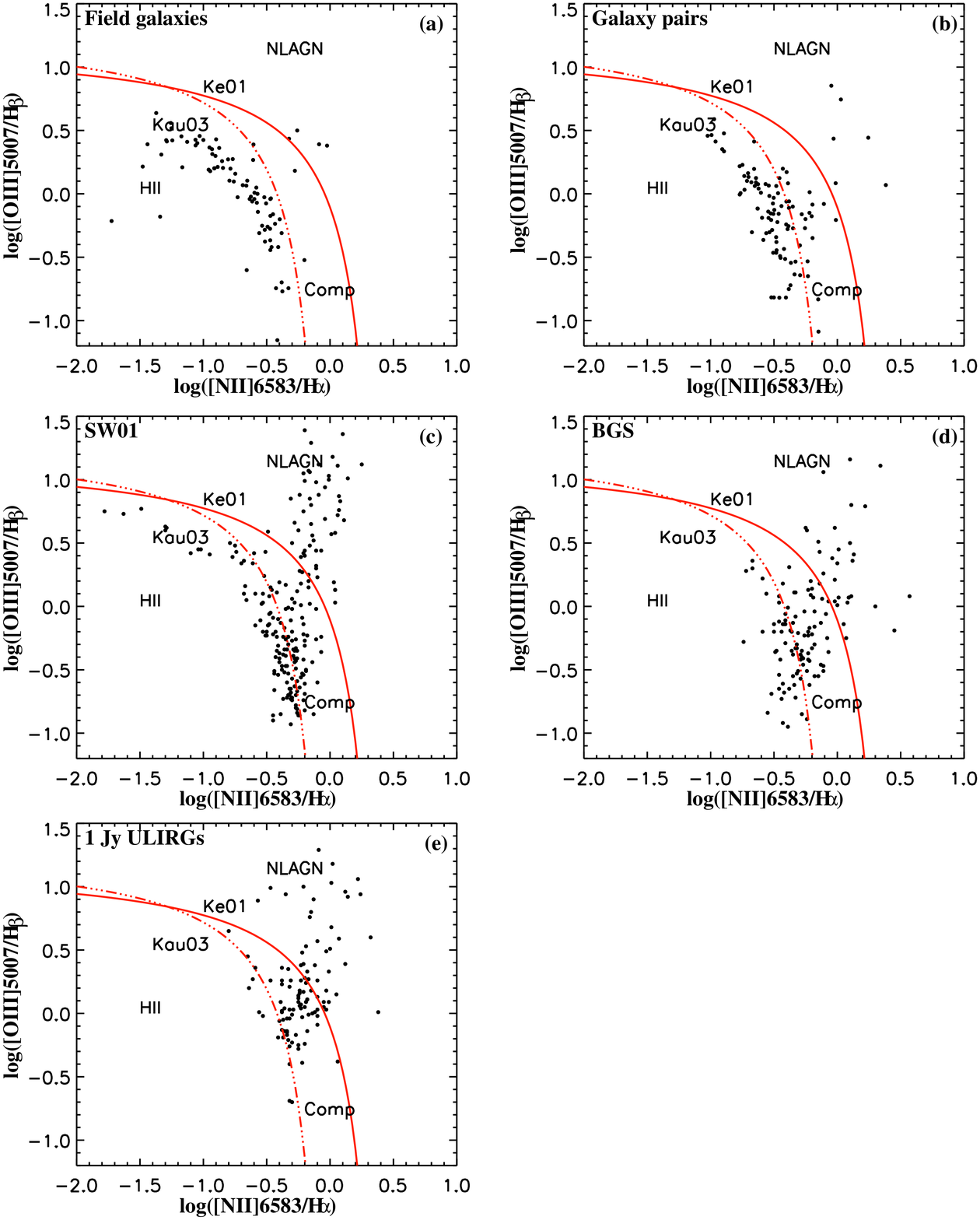}
\caption{   
Composite galaxies on the ``\nii~ diagram" for different types of galaxy samples. 
(a) Field galaxy sample: NFGS \citep{jansen00},
(b) Galaxy pair sample: BGK00 \citep{barton00},
(c) SW01 sample,
(d) BGS sample, 
(e) 1{\ts}Jy ULIRG sample.
}
\label{fig:sampleall}
\end{figure*}

In Fig.~\ref{fig:sampleall}, we show 5 different galaxy samples on the \nii/H$\alpha$ versus \oiii/H$\beta$ diagnostic diagram.  
The new Ke06 classification boundaries are indicated in red.   We supplement the 1{\ts}Jy ULIRG, BGS and SW01 samples with 
two optically selected samples: the NFGS field galaxy sample \citep{jansen00} and the BGK00 galaxy close pair sample 
\citep{barton00}.  The samples in Fig.~\ref{fig:sampleall} are ordered from (a) to (e) by the level of interaction activity $-$ from 
no interaction (NFGS field galaxies), to non-IR selected galaxy pairs, to luminous IR-selected galaxies, 
and to the most luminous IR-selected galaxies (1{\ts}Jy ULIRGs).  

The fraction of starburst-AGN composite galaxies increases from the field galaxy sample (3.6\%) $\rightarrow$ galaxy pair 
sample (22.8\% ) $\rightarrow$ SW01 galaxies (29.6\%) $\rightarrow$ BGS galaxies (dominated by LIRGs) (37\%) 
$\rightarrow$ 1{\ts}Jy ULIRGs (49.1\%).  This plot indicates that the stronger the interaction between galaxies is, the more 
likely they are to contain starburst-AGN composite galaxies.

Our new results offer some hope for resolving previous disputes concerning the evolution of starburst  
and AGN activity in gas-rich major mergers, in particular for ULIRGs.  We have shown that a large fraction of ULIRGs are of composite starburst-AGN spectral type, with line ratios (or $D_{\rm AGN}$) indicating an intermediate class between pure starbursts and pure AGN.   Spectral classification as a function of merger evolution does not change abruptly from ``pure" 
starburst into ``pure" AGN.  ULIRGs may spend a fairly large fraction of their merger history in a phase where starburst and AGN both make significant contributions to the EUV radiation.   In luminous IR mergers, starburst-AGN composites appear to 
``bridge" the spectral evolution from pure starburst to AGN-dominated activity as the merger progresses.  
For ULIRGs, we have shown that an initial apparent decrease in starburst activity from the wide binary to close binary 
stage is accompanied by a rise in the starburst-AGN composites.   Similarly, as the merger reaches its final stages, the fall in starburst-AGN composites is followed by a rise in Seyfert activity.  These effects can be seen in Fig.~\ref{fig:moru}, where the starburst-AGN composite activity peaks in the diffuse merger stage.  We do not have sufficient data to determine whether a 
 similar scenario occurs at the lower IR luminosities spanned by the BGS and SW01 samples ($L_{IR}<10^{12}L_{\odot}$).

\subsection{The Merger Scenario for LIRGs $\rightarrow$ ULIRGs}
There is now substantial agreement that ULIRGs are triggered by major
mergers of gas-rich spirals \citep[see the review by ][]{sanders96}.  In
this merger scenario, the LIRG phase begins when tidal interactions 
between merging gas-rich disk galaxies initially trigger wide-spread
starburst activity in one or both disks.  As the merger progresses, gas is
funneled towards the merger nucleus and activates nuclear starbursts and  
AGN, further increasing the IR luminosity.  However, many questions still remain.  
The precise link between AGN fueling, the gas dynamics, and star formation 
either on a nuclear or a global scale is still poorly determined.  The relative strength 
of starburst and AGN activity, particularly in the ULIRGs is still debated, and the
 question of whether the majority of LIRG mergers become ULIRGs has not been 
 clearly addressed.

Our new results shown in Figs. 9-16 provide new insight into how starburst and AGN activity evolve during interaction/merger of IR-luminous galaxies.

For lower luminosity IR-galaxies ($L_{\rm IR} < 10^{12} L_{\odot}$), Seyferts are rare ($D_{\rm AGN} = 0.25-0.35$), and starbursts strongly contribute to the EUV radiation field either as ``pure" starbursts or as composites.  The majority of lower luminosity LIRGs are wide or close pairs.  These pairs may either evolve into the single merger stage without becoming ULIRGs, or they may enter the ULIRG phase during or shortly after the close pair stage.  We consider both cases:

(1) Some lower luminosity IR galaxies ($L_{\rm IR} < 10^{12} L_{\odot}$) may reach the final single merger stage without becoming ULIRGs.  As these non-ULIRG galaxies evolve from the close pair to the single merger stage, the mean IR luminosity increases by $\sim 2\times$ (to $L_{\rm IR} \sim 10^11.5 L_{\odot}$), with only a relatively small rise in the mean 
value of $D_{\rm AGN}$($\sim$0.4).  These results suggest that non-ULIRGs undergo a rise in starburst activity but only modest AGN growth as they evolve from the close pair to the single merger stage.  Some of these lower luminosity single mergers may be the product of minor mergers of objects with mass ratios larger than 5:1 \citep{ishida04, ishida07}.

(2) The lower luminosity pairs that may become ULIRGs are likely to be major mergers with mass ratios closer to unity \citep{ishida04, ishida07}.  In the ULIRG phase, the AGN contribution is elevated at all merger stages.  We observe clear changes in the optical classification of ULIRGs as a function of merger stage.  The diffuse merger stage (compared to the close binary stage) shows a dramatic increase in starburst-AGN composites (from $\sim$45\% to $\sim$80\%).  In the subsequent compact/old merger stages, the fraction of composite galaxies falls, and AGN dominate.  The mean value of $D_{\rm AGN}$ rises (by $\sim$0.22{\ts}dex $-$ from 0.43 to 0.65), and the mean value of $L_{\rm IR}$ rises ($\sim 1.7\times$), along with the emergence of a substantial Seyfert{\ts}1 fraction ($\sim$25\%).   These results suggest that during the diffuse merger stage, the AGN becomes more powerful and increasingly visible via the optical emission-line ratios.   Once the merged nucleus begins forming a core (compact merger stage) starburst activity may be subsiding (and possible dust obscuration clears) as AGN activity becomes prominent.  We tentatively suggest that the dramatic increase in $D_{\rm AGN}$ and the emergence of a substantial population of Seyfert{\ts}1s (accompanied by a $\sim 2\times$ increase in $L_{\rm IR}$) may signify a significant ``blowout" phase, as ULIRGs transition to optical QSOs \citep[e.g., ][]{hopkins05}.  A larger sample of ULIRGs is required to test this idea.

\section{SUMMARY}\label{summary}
We apply the new SDSS semi-empirical optical spectral classification scheme to three IR-selected galaxy 
samples $-$ the 1{\ts}Jy ULIRG Sample, the {\it IRAS} Bright Galaxy Sample and the Southern Warm 
IR Galaxy Sample.  Because the new classification scheme is substantially different from previous methods, 
the new scheme is used to yield insights into the relationship between starburst and AGN activity and 
merger progress.  We utilize optical and near-IR images to determine the projected separation between 
pairs in our samples, and to classify the galaxies morphologically.  The projected separation and 
morphological classification are used as relative tracers of merger progress.  For each sample, we investigate 
how the optical classification and the relative AGN contribution, $D_{\rm AGN}$, changes as a  function 
of  $L_{\rm IR}$ and merger  progress.   

We find that: 

1.  IR-luminous LINERs are rare; There are very few LINERs in the 1{\ts}Jy ULIRG, and the BGS  and SW01 samples.  The rarity of LINERs in the 1{\ts}Jy ULIRG and SW01 samples may be 
at least partly due to selection effects. Nearly all of the previously classified IR-luminous LINERs are starburst/AGN  composite galaxies in the Ke06 classification scheme. 
 
2.  The new classification scheme reveals a clear evolutionary scenario for ULIRGs from starburst-driven activity in the early merger stages, composite starburst-AGN activity intermediate merger stages to AGN-dominated emission at late merger stages.  The fraction of composite galaxies  rises from 45\% to 80\% between the wide binary and diffuse merger stages and appears to ``bridge" pure starburst and Seyfert galaxies.   Galaxies at the diffuse merger stage are key for  future 
investigations into the relationship between starburst and AGN activity in ULIRGs.

3. We find that advanced mergers preferentially occur in ULIRG samples.  We suggest that the transition into the ULIRG phase occurs close to or during the diffuse merger stage in which the nuclei of the two merging galaxies are coalescing.  

4.  At later merger stages in ULIRGs, when the single nucleus is forming a core, the fraction of pure-Seyfert objects rises 
dramatically.  This stage corresponds to the highest  $L_{\rm IR}$ in the 1{\ts}Jy ULIRG sample.  At this stage, 
we propose that\ (a) starburst activity subsides, allowing the AGN to dominate the energy budget, and/or\  (b) dust 
obscuration surrounding the AGN clears, allowing the AGN radiation field to ionize the surrounding gas, and/or (c) AGN 
activity increases.

5.  Seyfert{\ts}1s appear to occur only in the final ``compact/old merging" stages of ULIRGs.  A larger sample of Seyfert{\ts}1s is required to determine the significance of this result.   If this result holds for larger samples, we hypothesize that a rise in Seyfert 1 galaxies at late merger stages may signify a significant ``blowout" phase, as ULIRGs transition to optical QSOs.

6.  There is no significant change in spectral types for the non-ULIRG BGS and SW01 samples.  

Understanding the behavior of composite galaxies may help to build a more 
concrete picture of the merger process for all IR-luminous galaxies.  
Our future work includes integral field spectroscopy of composite galaxies and a detailed comparison between our results and
 the evolutionary merger models from  numerical simulations
\citet{barnes04, hopkins05,hopkins06a, hopkins06b, hopkins07a, hopkins07b}.

\acknowledgments
We thank D.-C. Kim for providing the images for the 1{\ts}Jy ULIRG sample and 
Sturm, E. for his mid-infrared data.  We also thank T. Heckman for helpful 
comments. We are grateful to the referee for some very good suggestions/comments that significantly improved this paper. 
This work has made use of the Digitized Sky Surveys (DSS) that were produced at the 
Space Telescope Science Institute, under US government 
grant NAG W-2166, and the NASA/IPAC Extragalactic Database (NED), which is operated by the 
Jet Propulsion Laboratory, California Institute of Technology.  The 2MASS data were obtained from the 
NASA/IPAC Infrared Science Archive, which is operated by the Jet Propulsion Laboratory, 
California Institute of Technology, under contract with NASA.

\clearpage
\begin{appendix}
\section{$D_{\rm AGN}$ from different BPT diagrams}\label{defdagn}

Theoretical models have shown that the distribution of galaxies on the BPT diagrams are mainly driven by 
variations of parameters such as  metallicity, stellar age, ISM pressure and ionization parameter \citep{dopita00,kewley01,groves04,dopita06}.  
The different sensitivity to these parameters  of the four line ratios makes the three BPT diagrams distinct in distinguishing
 certain branches of galaxies.  
 
 \citet{kewley06} defined $D_{\rm AGN}$ on the \oi/H$\alpha$ diagram because the Seyfert and LINER 
 branches are most clearly separated on the  \oi/H$\alpha$ diagram.  In this case, $D_{\rm AGN}$  can be defined 
 separately for Seyferts and LINERs. The disadvantage of this diagram is that pure star-forming galaxies 
have to be removed first using the other two diagrams, because star-forming galaxies and starburst-AGN composites 
occupy similar regions on the \oi/H$\alpha$ diagram. 
 
\begin{figure*}[!ht]
\epsscale{0.8}
\plotone{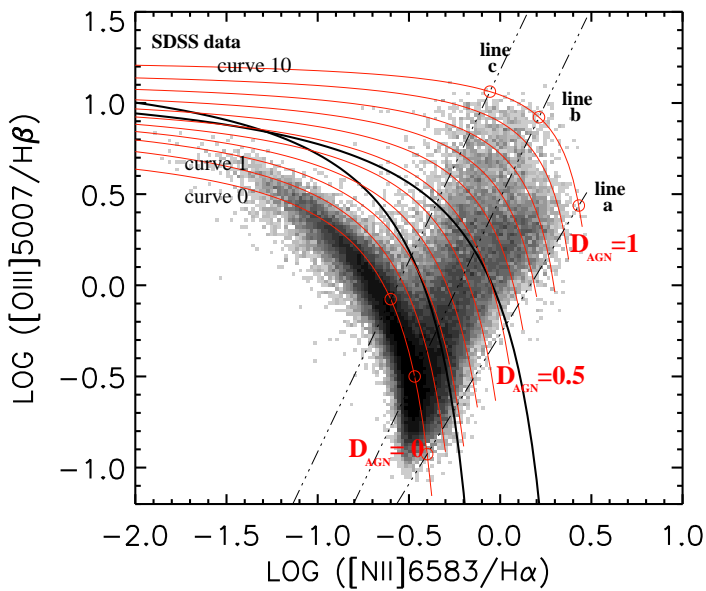}
\caption{ 
$D_{\rm AGN}$ defined on the ``\nii" diagram.  ``Base" and ``peak" points are chosen to fit the outer boundary 
of the star-forming sequence and the $D_{\rm AGN}=1$ curve. The regions between the boundary and curve are then 
divided into 10 bins to define the value of $D_{\rm AGN}$. 
}
\label{fig:sample9}
\end{figure*}

The advantage of the  \nii/H$\alpha$ diagram is that the star-forming sequence is most tightly formed, 
giving a better contrast with the starburst-AGN branch. Also,  \nii~has higher signal-to-noise than \oi. 
However, in this diagram, Seyferts and LINERs can not be well separated.   We define $D_{\rm AGN}$ on the  
\nii/H$\alpha$ diagram in Fig.~\ref{fig:sample9}: first, we empirically fit the SDSS star-forming sequence with a curve. 
Then, three points are chosen from the curve as ``base points" (empty red circles in the lower left region in Fig.~\ref{fig:sample9}).   
Another three ``peak points" (empty red circles in the upper right region in Fig.~\ref{fig:sample9}) are chosen from the 
upper right region of the diagram as $D_{\rm AGN}=1$.  These points construct three ``evolutionary line" intervals 
(line a, b, c in  Fig.~\ref{fig:sample9}).  We divide the intervals equally into 10 bins which define regions for different 
values of $D_{\rm AGN}$, e.g., for the region below curve 0, $D_{\rm AGN}=$0 ; for the region above curve 0 and 
below curve 1, $D_{\rm AGN}=$0.1, ..., and $D_{\rm AGN}=$ 1 for the region above curve 10.   Note that $D_{\rm AGN}$ 
is defined differently from the ``radial-arc" system used in \citet{kewley06}, specifically for the \oi/H$\alpha$ diagram, 
because the starburst-AGN branch does not simply develop from one base point on the  \nii/H$\alpha$ diagram. 
 
\begin{figure*}[!ht]
\epsscale{1.2}
\plotone{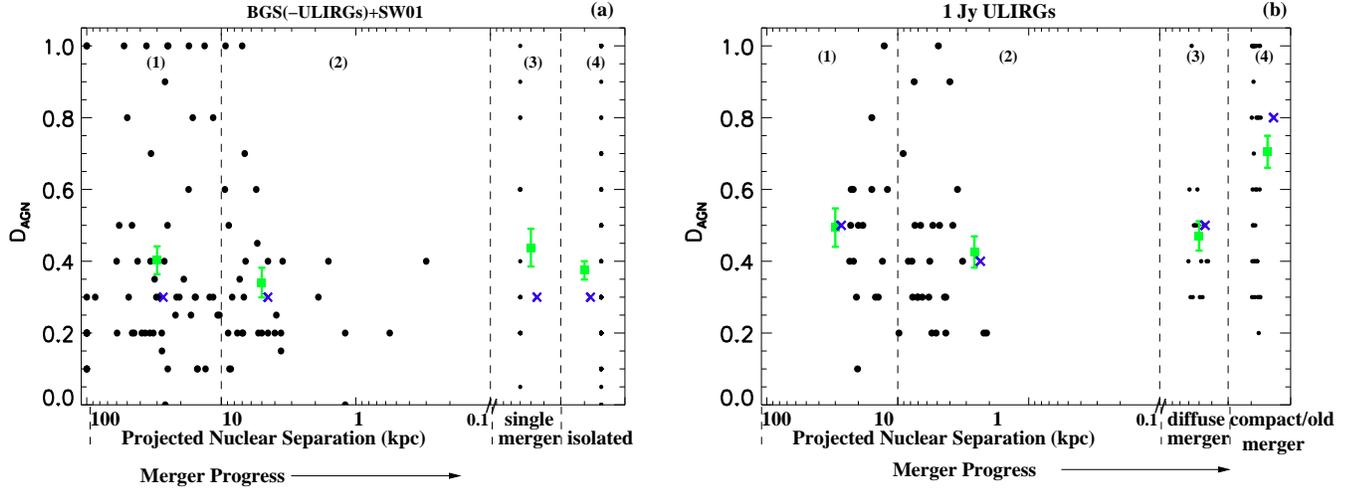}
\caption{ As in Fig.~\ref{fig:lirdagn_bgsw} and Fig.~\ref{fig:lirdagn_ulig},  but using $D_{\rm AGN}$ as defined on the  \nii/H$\alpha$ diagram. }
\label{fig:mordagn2}
\end{figure*}

We list the $D_{\rm AGN}$ defined on both diagrams in Tables~\ref{tb1}-\ref{tb3}.  There is little difference in 
the two differently defined $D_{\rm AGN}$, however, it is not meaningful to compare the absolute value of $D_{\rm AGN}$.
As emphasized in the text, only the relative value of $D_{\rm AGN}$ is useful.  As an example, we show in Fig.~\ref{fig:mordagn2} 
that our results in \S~\ref{dagn_ns} do not change when we apply the $D_{\rm AGN}$ defined on the  \nii/H$\alpha$ 
diagram.   Defining $D_{\rm AGN}$ on the \sii/H$\alpha$ diagram does not change in the results.
 
There is a good relation between $D_{\rm AGN}$ and spectral types:  the mean value of $D_{\rm AGN}$ increases from 
star-forming galaxies, to starburst-AGN composites, and to Seyfert{\ts}2 and LINERs.  Table~\ref{tb10} lists the spectral types 
and their corresponding mean and median $D_{\rm AGN}$.  The values are based on $\sim400$ galaxies from the 
1{\ts}Jy ULIRG, BGS and SW01 samples, as given in Table~\ref{tb1},~\ref{tb2} and ~\ref{tb3}.  Star-forming galaxies have 
$D_{\rm AGN} \leq 0.2$,  starburst-AGN composites have a mean $D_{\rm AGN}=0.4(0.4) $,  and Seyferts/LINERs have 
$D_{\rm AGN}=0.7(0.8)$.   We assign $D_{\rm AGN} = 1$ to Seyfert{\ts}1 objects since they are not included in 
the BPT diagrams and are almost certainly dominated by an AGN.


\section{Classification criteria and ``ambiguous" class}\label{ambissue}

In \S~\ref{class}, we use the 2-of-3 criterion to assign a class to our samples, based on agreement between two out of the
three standard optical diagnostic diagrams.  The 2-of-3 criterion has two advantages: (1) the final classification is less 
sensitive to low signal-to-noise in either \oi\ or \sii\, which may sometimes be weak, and (2) galaxies without \sii\ or \oi\ measurements can be assigned a class based on the remaining one or two diagnostic diagrams.  

\begin{figure*}[!ht]
\epsscale{1.34}
\plotone{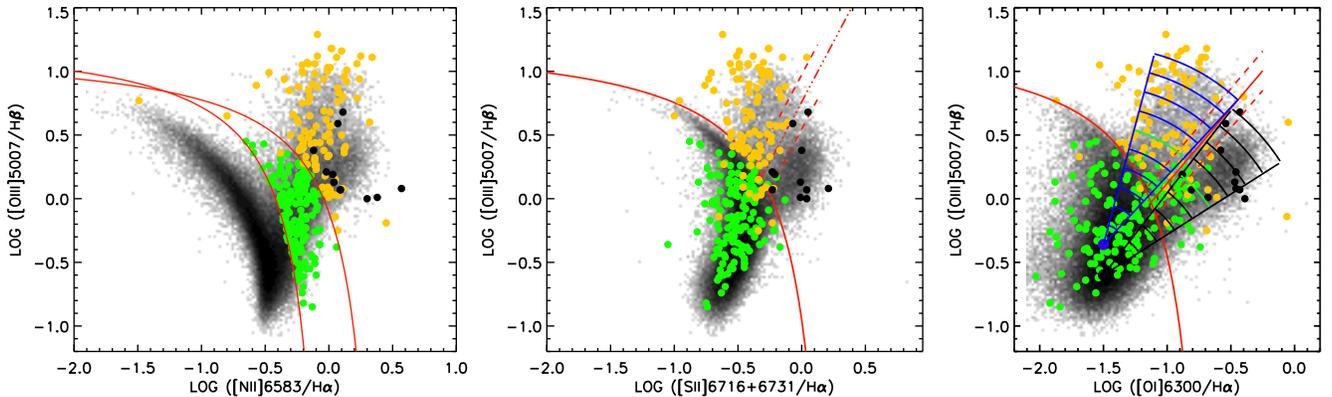}
\caption{Classified galaxies in the 3 samples using the 2-of-3 criterion. Green: composites. Orange: Seyfert{\it}2, 
Black: LINERs. Solid red lines are the new classification boundaries as in Fig.~\ref{fig:class}. 
Wedges on the \oi/H$\alpha$ diagram are used to derive $D_{\rm AGN}$ as described in Fig.~\ref{fig:DAGN}}
\label{fig:bptamb}
\end{figure*}

In Figure~\ref{fig:bptamb}, we show the galaxies in the 3 samples classified using the 2-of-3 criterion on the BPT diagrams.  As outlined in \citet{kewley06}, the \NIIHa\ line ratio is more sensitive to the presence of low-level AGN than \SIIHa\ or \OIHa because \NIIHa\ is more 
sensitive to metallicity. The log(\NIIHa) line ratio is a linear function of the nebular metallicity until high metallicities where the log(\NIIHa) ratio saturates. This saturation point causes the star-forming sequence to be almost vertical at log$($\NIIHa$)=0.5$. Any AGN contribution moves 
 \NIIHa\ above this saturation level, allowing the identification of galaxies with even small AGN contributions.   
 
 The \SIIHa\ and \OIHa\ line ratios can not be used to distinguish composite galaxies from pure star-forming galaxies.
In the \SIIHa\ and \OIHa\ diagnostic diagrams, composite galaxies occupy a region that is degenerate with the star-forming galaxy sequence.  This problem is caused by the relationship 
between \SIIHa\ and \OIHa\ and metallicity.   The AGN-starburst mixing sequence begins at the high metallicity end of the star-forming galaxy sequence.  The \SIIHa\ and \OIHa\ line ratios 
(unlike the  \NIIHa\ ratio) are double-valued with metallicity over the range of observed \SIIHa\ and \OIHa\ ratios in nearby galaxies.  The highest metallicity star-forming galaxies occur at
low  \SIIHa\ and \OIHa\ ratios (log$($\SIIHa$)= -0.6$ to $-0.9$, and log$($\OIHa$)= -2.0$ to $-1.4$)
\citep[see Figures in][]{dopita00,kewley01}.   Therefore, a large AGN contribution 
($\sim 40-50$\%) is required for a composite galaxy to rise above the star-forming galaxy sequence.  

The 3-of-3 criterion is a more stringent method of classification that is based on all three diagnostic diagrams.  This method allows
ambiguous galaxies to be classified as those galaxies that have one classification in one or two diagrams and  a different classification in the remaining diagram(s).  The 3-of-3 criterion is suitable for galaxies with high S/N spectra where all
five diagnostic emission-lines have $S/N>3$ \citep[{\it e.g.,}][]{kewley06}.  Samples with lower S/N emission-lines may contain a large fraction of ambiguous galaxies if  the 3-of-3 criterion is applied.  

According to the  3-of-3 scheme, the 1{\ts}Jy ULIRGs contain 4 (4.0\%) HII-region galaxies, 18 (18\%) starburst-AGN composites, 23 (23\%) Seyfert{\ts}2, 10 (10\%) Seyfert{\ts}1, and 3 (3\%) LINERs. The remaining 41 (41\%) galaxies are ambiguous, illustrating 
the effect of low S/N for the \sii\ and/or \oi\ emission-lines in this sample.  

For the LIRGs in the BGS sample, a total of 74 single nucleus galaxies have measured spectra on all three diagrams. 
Using the 3-of-3 criterion, 22 (29.7\%) are HII-region galaxies, 19 (25.7\%) are starburst-AGN composites, 8 (10.8\%) are Seyfert{\ts}2, 1 galaxy (1.4\%) is a Seyfert{\ts}1, and 5 (6.7\%) are LINERs.   The remaining 19 (25.7\%) galaxies are ambiguous.  

For the SW01 sample, a total of 175 single nucleus objects that have measured emission line fluxes with $S/N > 3 \sigma$.  The majority of the sample has $S/N > 8 \sigma$ for all five diagnostic lines.  In the 3-of-3 scheme, we obtain 70 (40.0\%) HII-region galaxies, 48 (27.4\%) starburst-AGN composites,  33 (18.9\%) Seyfert{\ts}2, 10 (5.7\%) Seyfert{\ts}1, and 2 (1.1\%) LINERs.   The remaining 12 (6.8\%) galaxies are ambiguous.  

With the 3-of-3 criterion,we obtain a large fraction of ambiguous galaxies in the ULIRGs and BGS samples (41\% and 25.7\% respectively), while the fraction of ambiguous galaxies in the SW01 sample is relatively small (7\%).   The fraction of ambiguous 
galaxies in the SW01 sample is consistent with the fraction of ambiguous galaxies in the Sloan Digital Sky Survey where a $S/N>3$ cut has been applied. 

Among the ambiguous galaxies in the ULIRGs, 68\% are ambiguous between composites and Seyferts/LINERs, i.e., 
they lie within the composite galaxy region on the \nii/H$\alpha$ diagram, but lie above the Ke01 line
on either (or both) the \sii/H$\alpha$ or the \oi/H$\alpha$ diagram.   Many of these galaxies lie within the $\pm 0.1$ dex error region of either (or both) the \sii/H$\alpha$ or the \oi/H$\alpha$ classification line and hence their classification with these diagram(s) is uncertain.    \citet{hill99} and \citet{hill01} investigate
ambiguous galaxies using near-infrared spectroscopy and radio observations.  They conclude that ambiguous galaxies are 
 starburst-AGN composites.  Spitzer spectroscopy of our ambiguous 1 Jy ULIRGs supports this conclusion; 5/6 of the 
ULIRG ambiguous galaxies observed by \citet{ima07} have evidence (either strong or 
tentative) of a buried AGN.   Our results are unchanged if the 3-of-3 criterion is applied and if the ambiguous 
galaxies found using this criterion are starburst-AGN composites.

\section{Double Nuclei Galaxies with Different Spectral Types}\label{dnissue}

There are 23 double nuclei galaxies in the BGS. Of these, 13 have the same spectral type for
the two nuclei (7 HII, 6 composites). For the remaining 10 galaxies, 5 have classes available only for one nucleus (4 HII, 1 composite); the remaining 5 have different spectral types for each nucleus  (1 composite/HII, 2 composite/Seyfert{\ts}2, and 2 HII/Seyfert{\ts}2). 
 There are 9 double nuclei galaxies in the 1Jy sample.  A total of  4 of these double nuclei galaxies  have the same spectral type for each nucleus (composite).  The remaining 5/ 9 1 Jy ULIRGs all have a composite class for one nucleus (3 composite/HII, 2 composite/Seyfert{\ts}2).
 There are 17 double nuclei galaxies in the SW01 sample.  A total of 13 of these double nuclei galaxies have the same spectral type (8 HII, 3 composite, 2 Seyfert{\ts}2).
   Only 4 of the 17 double nuclei galaxies in SW01 have different spectral types (1 composite/LINER, 2 composite/HII, and 1 HII/Seyfert{\ts}2). 
     
There are only 14 objects across all three samples that have different spectral types for each nucleus.   
We estimate the ranges of uncertainty in our composite classification by first assigning all the double nuclei galaxies with different spectral types as composites in order to derive an upper limit for the ``composite" fraction.   We next assign all of the double nuclei galaxies with different spectral types the alternative types listed above to derive a lower limit of the ``composite" fraction. 
 The change in the fraction (shown as dashed lines in Figures~\ref{fig:moru} - \ref{fig:moru2}) of the composite galaxies indicates the range of uncertainty 
 introduced by these 14 objects.
 \end{appendix}

\clearpage
{}

\clearpage
\LongTables

\begin{turnpage}
\begin{deluxetable}{lcccccccc}
\tabletypesize{\scriptsize}
\tablenum{1}
\tablecolumns{9}
\tablewidth{0pt}
\tablecaption{Spectral Classification Result for the 1{\ts}Jy ULIRG Sample\label{tb1}}
\tablehead{
\colhead{Name} &
\multicolumn{5}{c}{Spectral Type}&
\colhead{$D_{\rm AGN}$} &
\colhead{Morphology} &
\colhead{$log(L_{\rm IR}/L_{\odot})$}\\
\cline{2-6}
\colhead{IRAS FSC} &
\multicolumn{1}{c}{[NII]} &
\multicolumn{1}{c}{[SII]} &
\multicolumn{1}{c}{[OI]} &
\multicolumn{1}{c}{Adopt}& 
\multicolumn{1}{c}{``VO87"} &
\colhead{}&
\colhead{Class} &
\colhead{}\\
\colhead{(1)} &
\multicolumn{1}{c}{(2)}& 
\multicolumn{1}{c}{(3)} &
\multicolumn{1}{c}{(4)}& 
\multicolumn{1}{c}{(5)} &
\multicolumn{1}{c}{(6)}& 
\multicolumn{1}{c}{(7)}& 
\colhead{(8)}&
\colhead{(9)}\\
}
\startdata
  00091$-$0738           &cp     &H &H         &cp  &H: &0.1 (0.2) &b &12.19 \\
  00188$-$0856           &S2      &H &S2     &S2:  &L&0.7(0.8)   &e &12.33 \\
  00397$-$1312           &cp       &H &H        &cp  &H&0.4   (0.3) &e & 12.90\\
  00456$-$2904           &cp       &H &H       &cp   &H&0.2   (0.3)& a& 12.12\\
  00482$-$2721           &cp       &H: &L       &cp:  &L:&0.8   (0.5)& b& 12.00\\
  01004$-$2237           &cp       &S2&S2     &S2: &H:&0.7    (0.4)&e & 12.24\\
  01166$-$0844           &H::   &H &\nodata &H: &H:&0 (0.2)   & b&12.03 \\
  01199$-$2307           &H:       &H: &S2:    &H:  &H:& 0.6   (1)&a &12.26 \\
\enddata
\tablecomments{Col. (1): Galaxy Name. Col. (2) to Col. (4)-Optical spectral types derived in this work from the 
\nii~$\lambda6583$, \sii~$\lambda \lambda6716, 6731$, and \oi~$\lambda6300$ diagrams respectively. 
The typical uncertainty on the emission-line ratios is $\sim$10\%; colons in Col. (3) and Col. (4) indicate larger
uncertainty of the line ratios of $\sim$25\% for  \sii/H$\alpha$ and \oi/H$\alpha$, respectively;
double colons in Col. (2) indicate larger uncertainty of the line ratios of $\sim$25\% for  \oiii/H$\beta$.
Col. (5) is the adopted 
spectral type using the 2-of-3 scheme; single colons indicate the object does not have the same spectral type in all 3
diagrams, i.e., the ambiguous class. Double colons means the classification does not agree on any of the diagrams, 
 the classification in those rare cases is highly uncertain. 
 For comparison, Col. (6) lists the VO87 classification results \citep{veilleux99}.
Symbols: H = star-forming galaxies, L = LINERs, S2 = Seyfert{\ts}2 galaxies, S1 = Seyfert{\ts}1 galaxies,
cp = starburst-AGN composite galaxies. 
Col. (7): $D_{\rm AGN}$ defined on the \oi/H$\alpha$ diagram, for comparison, in the bracket we also 
list the $D_{\rm AGN}$ derived from the \nii/H$\alpha$ diagram (see Appendix~\ref{defdagn}).
Col. (8): Morphology class as adopted from \citet{veilleux02}. a: wide binary; b: close binary; c: diffuse merger; d: compact merger;
e: old merger; tpl: triple mergers; iso: isolated. 
Col.(9): $log(L_{\rm IR}/L_{\odot})$, data from \citet{veilleux99}. }
\end{deluxetable}
\end{turnpage}
\clearpage

\LongTables
\begin{turnpage}
\begin{deluxetable}{lcccccccccc}
\tabletypesize{\scriptsize}
\tablewidth{0pt}
\tablenum{2}
\tablecolumns{11}
\tablecaption{ Spectral Classification Result and Morphology 
 for the IRAS BGS Sample\label{tb2}}
\tablehead{
\colhead{Name} &
\multicolumn{5}{c}{Spectral Type}&
\colhead{$D_{\rm AGN}$}&
\colhead{{\it ns }}&
\colhead{``reference"}&
\colhead{Morphology}&
\colhead{$log(L_{\rm IR}/L_{\odot})$}\\
\cline{2-6}
\colhead{} &
\multicolumn{1}{c}{[NII]} &
\multicolumn{1}{c}{[SII]} &
\multicolumn{1}{c}{[OI]} &
\multicolumn{1}{c}{Adopt} &
\multicolumn{1}{c}{``VO87"}&
\colhead{}&
\colhead{kpc}&
\colhead{}&
\colhead{Class}&
\colhead{}\\ 
\colhead{(1)} &
\multicolumn{1}{c}{(2)} &
\multicolumn{1}{c}{(3)} &
\multicolumn{1}{c}{(4)} &
\multicolumn{1}{c}{(5)} &
\multicolumn{1}{c}{(6)}&
\colhead{(7)}&
\colhead{(8)}&
\colhead{(9)}&
\colhead{(10)}&
\colhead{(11)}\\
}
\startdata
NGC 23             &cp &H &H                 &cp&H&0.2(0.3)  &-2 &  2MASS, DSS &iso & 11.05\\
NGC 34             & S2& S2&S2             &S2&S2&0.7(0.8)  & -1 &  2MASS, DSS &cde &11.41 \\
MCG-02-01-051:S    &H:: &H &H         &H  &H&0(0.2)  & 27.7 &  2MASS &a & 11.32\\
MCG-02-01-051:N    & H& H&H            &H&H&0(0.1)  &27.7  &  2MASS & a&\nodata \\
NGC 232            &cp &H &H                    &cp&H:& 0.4(0.5) &8.8  &  2MASS& b& 11.30\\
UGC 556            &cp &H &L                    &cp:&L&0.5(0.3)  & -2 &  2MASS, DSS &iso & 10.82\\
IC1623:N           &H &H &H                      &H &H&0(0.2)  &5.3  &  2MASS& b& 11.54\\ 
IC1623:SE          &H &H &H                    &H&H & 0(0.2) & 5.3 &  2MASS& b& \nodata\\
\enddata
\end{deluxetable}
\end{turnpage}

\LongTables
\begin{turnpage}
\begin{deluxetable}{lccccccccc}
\tabletypesize{\scriptsize}
\tablewidth{0pt}
\tablenum{3}
\tablecolumns{9}
\tablecaption{Spectral Classification Result 
and morphology for the SW01 Sample\label{tb3}}
\tablehead{
\colhead{Name} &
\multicolumn{4}{c}{Spectral Type}&
\colhead{$D_{\rm AGN}$}&
\colhead{{\it ns }}&
\colhead{``reference"}&
\colhead{Morphology}&
\colhead{$log(L_{\rm IR}/L_{\odot})$}\\
\cline{2-5}
\colhead{} &
\multicolumn{1}{c}{[NII]} &
\multicolumn{1}{c}{[SII]} &
\multicolumn{1}{c}{[OI]} &
\multicolumn{1}{c}{Adopt} &
\colhead{}&
\colhead{kpc}&
\colhead{}&
\colhead{Class}&
\colhead{}\\
\colhead{(1)} &
\multicolumn{1}{c}{(2)} &
\multicolumn{1}{c}{(3)} &
\multicolumn{1}{c}{(4)} &
\multicolumn{1}{c}{(5)} &
\colhead{(6)}&
\colhead{(7)}&
\colhead{(8)}&
\colhead{(9)}&
\colhead{(10)}\\
}
\startdata
$\ast$IRAS00091$-$0738       &\nodata  &\nodata  &\nodata  & \nodata &\nodata& -1& 2MASS& cde& 12.16\\
IRAS00163$-$1039(Arp256)~obj1 & H&H &H       &H &    0(0.2) &27.6&2MASS   & a&11.40 \\
IRAS00163$-$1039(Arp256)~obj2& H&H &H      &H &    0(0.2) &27.6&2MASS   & a&11.40 \\
IRAS00198$-$7926       & S2& S2& S2&          S2&  0.6(0.6)&     17.5& 2MASS& a& 11.98\\
IRAS00247$-$0203(NGC118)      &\nodata &\nodata &\nodata & \nodata& \nodata&    -2  & 2MASS& iso&11.35\\
IRAS00317$-$2804(NGC150)      & cp&H &H     & cp& 0.4(0.4) &     -2& 2MASS&iso & 10.14\\
IRAS00335$-$2732       &cp &H &S2         & cp:& 0.5(0.4)&     6.6 & cb96& b& 11.93\\
IRAS00344$-$3349~obj1       & H&H &H      &H &0(0.3)&    1.4   & hv94&b & 11.18\\
\enddata
\tablecomments{We only list the spectral type results for galaxies having available line ratios
with $S/N>3\sigma$ in
all three BPT diagrams. The projected radius used for searching companions is $\sim 100${\ts}kpc. 
Col. (1) to Col. (6), the same as Table 2.
Col. (7), projected nuclear separation. 
-1: single nucleus with tidal features; -2: isolated systems 
``:": lower limit; ``m": multiple (more than two) interacting objects.
Col. (8), references used to obtain projected separation. 2MASS: 2MASS images ($K_{\rm s}$-band);
DSS: DSS images; 
cb96: \citet{clementsbaker96}; hv95: \citet{heislearvader95}; vva68: \citet{vva68}; ke01: \citet{kewley01a};
tec00: \citet{tecza00}. Col. (9), adopted morphology class, a: wide binary; b: close binary; cde: single merger (either diffuse merger, 
compact merger, or old merger); iso: isolated system; m: multiple mergers.
Symbols: H = star-forming galaxies, L = LINERs, S2 = Seyfert{\ts}2 galaxies, S1 = Seyfert{\ts}1 galaxies,
cp = starburst-AGN composite galaxies.
Col. (10): $log(L_{\rm IR}/L_{\odot})$, data from \citet{kewley01a}.
}
\end{deluxetable}
\end{turnpage}

\clearpage
\begin{deluxetable}{lcccc}
\tabletypesize{\scriptsize}
\tablewidth{0pt}
\tablenum{4}
\tablecolumns{5}
\tablecaption{Spectral Classification as a Function of $L_{\rm IR}$ for the 
1{\ts}Jy ULIRG + BGS Samples\label{tb4}}
\tablehead{
\colhead{} & \multicolumn{4}{c}{log{\ts}($L_{\rm IR} / L_{\odot}$)} \\
\cline{2-5}\\
\colhead{spectral type} &
\colhead{$<11$} &
\colhead{11--11.99} & 
\colhead{12--12.29} &
\colhead{12.3--12.8}}
\startdata
 new classification, excluding double nuclei objects, 3-of-3 criterion& & & \\
\cline{1-5}\\
Total number&22&51&69&31 \\
 & & \\
starburst                             &8(36.4\%)&12(23.5\%)&3(4.3\%)&1(3.2\%) \\
starburst-AGN composite&8(36.4\%)&19(37.3\%)&33(47.8\%)&12(38.7\%)\\
Seyfert{\ts}2                       &1(4.5\%)&7(13.7\%)&13(18.8\%)&8(25.8\%)\\
Seyfert{\ts}1                      &0(0\%)&1(2\%)&2(2.9\%)&8(25.8\%) \\
LINER                              &2(9.1\%)&3(5.9\%)&3(4.3\%)&0(0\%)\\
Ambiguous                  &3(13.6\%)&9(17.6\%)&15(21.7\%)&2(6.5\%)\\
\cline{1-5}\\
 new classification, excluding double nuclei objects, 2-of-3 criterion& & & \\
\cline{1-5}\\
Total number&22&51&69&31 \\
 & & \\
starburst                              &10(45.5\%)&12(23.5\%)&7(10.2\%)&1(3.2\%) \\
starburst-AGN composite &8(36.3\%)&19(37.3\%)&33(47.8\%)&12(38.7\%)\\
Seyfert{\ts}2                        &2(9.1\%)&16(31.3\%)&24(34.8\%)&10(32.3\%)\\
Seyfert{\ts}1                         &0(0\%)&1(2\%)&2(2.9\%)&8(25.8\%) \\
LINER                                   &2(9.1\%)&3(5.9\%)&3(4.3\%)&0(0\%)\\
\cline{1-5}\\
\enddata
\tablecomments{ 
Number of galaxies and their fraction (shown in brackets) in each spectral
type class. Top rows are results
 from the 3-of-3 criterion (not including galaxies with double nuclei); bottom rows are results from 
the 2-of-3 criterion (not including galaxies with double nuclei). 
}
\end{deluxetable}

\begin{deluxetable}{lcccc}
\tabletypesize{\scriptsize}
\tablewidth{0pt}
\tablenum{5}
\tablecolumns{5}
\tablecaption{Spectral Classification as a Function of $L_{\rm IR}$ 
for the SW01 Sample\label{tb5}}
\tablehead{
\colhead{} & 
\multicolumn{4}{c}{log{\ts}($L_{\rm IR} / L_{\odot}$) } \\
\cline{2-5}\\
\colhead{spectral type} &
\colhead{8.0--10.19} &
\colhead{10.2--10.69} &
\colhead{10.7--11.09} &
\colhead{11.1-12.2} 
}
\startdata
new classification, excluding double nuclei objects, 3-of-3 criterion& & & \\
\cline{1-5}\\
Total number&37&49&49&41\\
 & & \\
starburst (HII region)                   &25(67.6\%) &25(51.0\%) &15(30.6\%)&5(12.2\%) \\
starburst-AGN composite           &7(18.9\%)  &13(26.5\%) &19(38.8\%)&15(36.5\%)\\
Seyfert{\ts}2                    &4(10.8\%)  &8(16.3\%)  &7(14.3\%) &14(34.2\%)\\
Seyfert{\ts}1                    &0(0\%)    &2(4.1\%)  &3(6.1\%)  &5(12.2\%) \\
LINER                        &1(2.7\%)  &0(0\%)     &1(2.0\%)  &0(0.0\%)\\
Ambiguous                    &0(0\%)     &1(2.1\%)   &4(8.2\%)  &2(4.9\%)\\
\enddata
\tablecomments{Since the fraction of ambiguous galaxies in SW01 is small, we only list the stringent 3-of-3 criterion results here.
Detail classification for a specific ambiguous galaxy can be found in Table~\ref{tb3}.
}
\end{deluxetable}

\clearpage
\begin{deluxetable}{lccccc}
\tabletypesize{\scriptsize}
\tablewidth{0pt}
\tablenum{6}
\tablecolumns{6}
\tablecaption{Spectral type as a function of morphology for the 1{\ts}Jy ULIRG Sample\label{tb6}}
\tablehead{
\colhead{} & 
\multicolumn{5}{c}{Morphology Group\tablenotemark{a}}\\
\cline{2-6}\\
\colhead{Spectral type} &
\colhead{Wide binary} &
\colhead{Close binary} &
\colhead{Diffuse merger} &
\colhead{Compact merger} &
\colhead{Old merger} \\
}
\startdata
 & & \\
Total number&17&29&15&30&11 \\
 & & \\
starburst                  &2(11.8\%$\pm$8.3\%) &5(17.2\%$\pm$7.7\%) &0(0\%$\pm$0\%) &0(0\%$\pm$0\%)    &0(0\%$\pm$0\%)\\
starburst-AGN composite           &8(47\%$\pm$16.7\%) &16(55.2\%$\pm$13.7\%) &12(80.0\%$\pm$23.1\%)&11(36.6\%$\pm$11.1\%) &4(36.4\%$\pm$18.1\%)\\
LINER                        &0(0\%$\pm$0\%)    &2(6.9\%$\pm$4.9\%)  &1(6.7\%$\pm$6.7\%) &0(0\%$\pm$0\%)    &0(0\%$\pm$0\%)\\
Seyfert 2                    &7(41.2\%$\pm$15.6\%) &6(20.7\%$\pm$8.4\%) &2(13.3\%$\pm$9.4\%)&11(36.7\%$\pm$11.1\%)&5(45.4\%$\pm$20.3\%) \\
Seyfert 1                    &0(0\%$\pm$0\%)    &0(0\%$\pm$0\%)    &0(0\%$\pm$0\%)   &8(26.7\%$\pm$9.4\%) &2(18.2\%$\pm$12.9\%)\\
\enddata
\tablenotetext{a}{Notes: The 1 isolated object and 5 possible triple-system objects are not included here.
The mean and standard error of the fractions are shown in the brackets.}
\end{deluxetable}

\begin{deluxetable}{lcccc}
\tabletypesize{\scriptsize}
\tablewidth{0pt}
\tablenum{7}
\tablecolumns{5}
\tablecaption{Spectral type as a function of morphology for the LIRGs in the BGS sample\label{tb7}}
\tablehead{
\colhead{} & 
\multicolumn{4}{c}{Morphology Group\tablenotemark{a}} \\
\cline{2-5}\\
\colhead{Spectral type} &
\colhead{Wide binary} &
\colhead{Close binary} &
\colhead{Single merger} &
\colhead{Isolated} \\
}
\startdata
 & & \\
Total number&20&13&9&35 \\
 & & \\
starburst                 &8(40.0\%) &6(46.1\%) &4(44.5\%)  &12(34.3\%)\\
starburst-AGN composite           &9(45.0\%) &5(38.5\%) &3(33.3\%)  &15(42.9\%)\\
LINER                        &0(0\%)    &1(7.7\%)  &1(11.1\%)  &3(8.6\%)\\
Seyfert 2                    &2(10.0\%) &1(7.7\%)  &1(11.1\%)  &5(14.2\%) \\
Seyfert 1                    &1(5.0\%)  &0(0\%)    &0(0\%)     &0(0\%)\\
\enddata
\tablecomments{Only 77 objects with 
both available spectral type and morphology classification are included.}
\tablenotetext{a}{Notes: wide binary ($ns > 10${\ts}kpc); close binary ($ns < 10${\ts}kpc); 
single merger: merger in the final merging stage where the two nuclei have merged into
a single one; isolated non-merging systems.}
\end{deluxetable}

\begin{deluxetable}{lcccc}
\tabletypesize{\scriptsize}
\tablewidth{0pt}
\tablenum{8}
\tablecolumns{5}
\tablecaption{Spectral type as a function of morphology for the SW01 sample\label{tb8}}
\tablehead{
\colhead{} & 
\multicolumn{4}{c}{Morphology Group\tablenotemark{a}} \\
\cline{1-5}\\
\colhead{Spectral type} &
\colhead{Wide binary} &
\colhead{Close binary} &
\colhead{Single merger} &
\colhead{Isolated} \\
}
\startdata
 & & \\
Total number&41&24&18&124 \\
 & & \\
starburst (HII region)                   &15(36.6\%) &14(58.4\%) &5(33.3\%)     &55(46.2\%)\\
starburst-AGN composite           &14(34.2\%) &5(20.8\%) &5(33.3\%)  &37(31.1\%)\\
LINER                        &0(0\%)    &2(8.3\%)  &0(0\%)    &1(0.8\%)\\
Seyfert 2                    &9(21.9\%) &2(8.3\%) &5(33.3\%)  &22(18.5\%) \\
Seyfert 1                    &3(7.3\%)    &1(4.2\%)    &0(0\%)  &4(3.4\%)\\
\enddata
\tablecomments{Only 207 objects with 
both available spectral type and morphology classification are included.}
\tablenotetext{a}{wide binary ($ns > 10${\ts}kpc); close binary ($ns<10${\ts}kpc); 
single merger: merger in the final merging stage where the two nuclei have merged into
a single one; isolated: non-merging systems.}
\end{deluxetable}

\clearpage
\begin{deluxetable}{lcccc}
\tabletypesize{\scriptsize}
\tablewidth{0pt}
\tablenum{9}
\tablecaption{Properties of LINERs in the 3 samples\label{tb9}}
\tablehead{
\colhead{Name} & 
\colhead{$d_{\rm [OI]}$\tablenotemark{a}(dex)} & 
\colhead{$d_{\rm [SII]}$(dex)} & 
\colhead{log{\ts}$(L_{\rm IR} / L_{\odot})$} & 
\colhead{morphology}\\
}
\startdata
& &1{\ts}Jy ULIRGs&&\\
\cline{1-5}\\
IRAS14504-1958&0.05&0.02&12.12&\nodata \\
IRAS04074-2801&0.15&0.18&12.14&diffuse merger\\
IRAS15327+2340(Arp220)&0.31&0.34&12.17&close binary, $ns=0.${\ts}kpc\\
\cline{1-5}\\
&&SW01 &&\\
\\
\cline{1-5}\\
IRAS05497-0728(NGC2110)&0.07&0.08&10.08&\nodata \\
IRAS16504+0228(NGC6240)&0.4&0.28&10.86&close binary,$ns=0.74${\ts}kpc\\
IRAS16399-0937,object2&0.05&0.08&11.07&\nodata \\
\\
\cline{1-5}\\
&&BGS&&\\
\cline{1-5}\\
IRAS05187-1017 &0.17&0.12&11.23&isolated?\\
IRAS01364-1042&0.35&0.05&11.76&\nodata\\
IRAS03134-0236(NGC1266)&0.13&0.44&10.34&isolated?\\
IRAS22359-2606(ESO534-G009)&0.54&0.32&10.61&isolated?\\
\enddata
\tablecomments{General properties of the 9 LINERs classified by the new scheme in the three samples.  }
\tablenotetext{a}{index distance from the Seyfert-LINER classification line in 
the BPT \oi  diagram. Available morphology information comes from images in NED.}
\end{deluxetable}

\begin{deluxetable}{lcc}
\tabletypesize{\scriptsize}
\tablewidth{0pt}
\tablenum{10}
\tablecolumns{3}
\tablecaption{Spectral Type and $D_{\rm AGN}$\label{tb10}}
\tablehead{
\colhead{} & 
\multicolumn{2}{c}{$D_{\rm AGN}$} \\
\cline{2-3}\\
\colhead{Spectral type} &
\colhead{Mean} &
\colhead{Median} \\
}
\startdata
starburst (HII region)                   & 0.2(0.2)        &0(0.2)   \\
starburst-AGN composite        & 0.4(0.4)     &0.3(0.4)  \\
LINER                        &0.8(0.8)       &0.9(0.9)  \\
Seyfert{\ts}2                    &0.8(0.7)        &0.8(0.7) \\
\enddata
\tablecomments{Col. (1) Spectral Types defined using the Ke06
 semi-emipirical scheme; Col. (2) $D_{\rm AGN}$ median value; 
Col. (3) $D_{\rm AGN}$ mean value.  Based on 420 galaxies from the 1{\ts}Jy ULIRG, BGS and SW01 samples.
Seyfert{\ts}1 objects are set to have $D_{\rm AGN}\equiv1$. }

\end{deluxetable}
 
\clearpage
\end{document}